\documentclass[a4paper,fleqn,usenatbib]{mnras}
\usepackage{newtxtext,newtxmath}
\usepackage[T1]{fontenc}
\usepackage{ae,aecompl}

\usepackage{color}

\usepackage{mathtools} 
\usepackage{amsmath} 
\usepackage{amsfonts} 
\usepackage{mathrsfs} 
\usepackage{graphicx} 
\usepackage{float} 
\usepackage{caption} 
\usepackage{bigstrut} 
\usepackage{tabularx} 
\usepackage{cancel} 
\usepackage{soul}
\usepackage{upgreek}

\usepackage{dcolumn}
\usepackage{bm}
\usepackage{xcolor}
\usepackage{ulem}

\usepackage{slashed} 
\usepackage{epstopdf} 

\newcommand{\msout}[1]{\text{\sout{\ensuremath{#1}}}} 
  {\color{red}}%
  {}
\newcommand{\vecnabla}{\bm \nabla}
\newcommand{\vecxi}{\bm \xi}

\title[Unstable modes of
  hypermassive compact stars]{Unstable modes of
  hypermassive compact stars driven by viscosity and gravitational radiation}

\author[P. B. Rau and A. Sedrakian]{
Peter B. Rau$^{1}$\thanks{E-mail: prau@uw.edu (Corresponding author. Current address: Institute for Nuclear Theory, University of Washington, Seattle, WA 98195-4550, U.S.A.)}
and Armen Sedrakian$^{2,3}$\thanks{E-mail: sedrakian@fias.uni-frankfurt.de}
\\
$^{1}$Cornell Center for Astrophysics and Planetary Science and Department of Astronomy, Cornell University, Ithaca, NY 14850, U.S.A.
\\
$^{2}$Frankfurt  Institute  for  Advanced  Studies,  D-60438  Frankfurt  am  Main,  Germany 
\\
$^{3}$Institute  of  Theoretical  Physics,  University  of  Wroc\l{}aw,  50-204  Wroc\l{}aw,  Poland}

\date{Accepted XXXX. Received XXXX; in original form XXXX}

\pubyear{2021}

\begin{document}
\label{firstpage}
\pagerange{\pageref{firstpage}--\pageref{lastpage}}
\maketitle

\begin{abstract}
We study the oscillations modes of differential rotating remnants of binary neutron star inspirals by modeling them as incompressible Riemann ellipsoids parametrized by the ratio $f$ of their internal circulation to the rotation frequency. The effects of viscosity and gravitational wave radiation on the modes are studied and it is shown that these bodies exhibit generic instabilities towards gravitational wave radiation akin to the Chandrasekhar--Friedman--Schutz instabilities for uniformly rotating stars. The odd-parity modes are unstable for all values of $f$ (except for the spherical model) and deformations, whereas the even parity unstable modes appear only in highly eccentric ellipsoids. We quantify the modification of the modes with varying mass of the model and the magnitude of the viscosity. The modes are weakly dependent on the range of the masses relevant to the binary neutron star mergers. Large turbulent viscosity can lead to a suppression of the gravitational wave instabilities, whereas kinematical viscosity has a negligible influence on the modes and their damping timescales.
\end{abstract}


\maketitle

\section{Introduction}

Hypermassive neutron stars (HMNS) are one of the possible outcomes of binary neutron star
(BNS) mergers. They are characterized as having a mass greater than the maximum value for a uniformly rotating neutron star, but are supported against gravitational collapse by the differential motion of their interior fluids. Numerical simulations (for recent examples, see 
~\cite{Kastaun2015,Kastaun2017,Dietrich2017,Radice2018,Most2019a,HanauskeParticles2010004,Ruiz2020,Chaurasia2020}) show
that after a period of $\sim10$ ms of nonlinear evolution, post-merger objects settle into equilibria supported against collapse by differential rotation. These equilibria
can be highly eccentric and emit gravitational radiation that could be
detected by gravitational wave interferometers~\citep{Abbott2019}. Depending on its mass, an HMNS may eventually collapse to a black hole or evolve to a
supramassive, uniformly rotating, compact star. Supramassive
neutron stars (SMNS), which have masses exceeding the maximum value for a nonrotating star, evolve by losing angular momentum into stable neutron stars or collapse to black holes, depending on whether their lower-spin counterparts along the constant baryon mass sequence
belong to a stable or unstable branch.

Two merger events, GW170817~\citep{Abbott2017} and
GW190425~\citep{Abbott2021}, observed by the LIGO--Virgo
collaboration, have definite BNS origin. The combined masses of
merged stars are $2.74M_{\odot}$ and $3.4M_{\odot}$ respectively. Whether
or not these mergers resulted in a prompt collapse to a black hole or
led to the formation of an HMNS is still an open question, but the
electromagnetic follow-up emission to GW170817 suggests that a
short-lived HMNS was formed~\citep{Shibata2017,Margalit2017,Bauswein2017,Gill2019}. Given
the detection rate of compact binary coalescence of one per week at the
current LIGO--Virgo collaboration sensitivity, the prospect of
observing BNS mergers in the future is optimistic.

The Fourier analysis of the gravitational wave spectrum emitted by an
HMNS in numerical simulations shows clear peaks at frequencies in the
range 1--4 kHz~\citep{Bauswein2014,Takami2014,Stergioulas2011}. Their
physical origin is obscured by the complex fluid dynamics of HMNS in
this ``ring-down'' phase. These frequencies could be visible to advanced
LIGO if the BNS merger is close enough (at distances of  the 
order of 40 Mpc) and other more sensitive telescopes, such
as the Einstein Telescope~\citep{Maggiore2020} and the Cosmic
Explorer~\citep{Reitze2019} in a more distant future. Apart from the
detection perspective, understanding the oscillation spectrum
and the stability of HMNS is important for assessing their
lifetimes, mechanism of collapse, and spectrum of oscillations
on longer (of the order of 10--100 ms) time scales.
Because of the high computational cost of running
BNS simulations on such timescales and the complexity of implementing
viscosity in full-scale numerical simulations, studies of
quasinormal modes of HMNS using semi-analytic methods are useful both
for covering large parameter spaces as well as gaining insights in the
physics of the oscillations and instabilities.

{\it Uniformly} rotating gravitationally-bound stars (as first
demonstrated for ellipsoidal bodies by \cite{Chandrasekhar1969}, 
hereafter abbreviated as EFE) undergo secular instabilities induced by
viscosity~\citep{Roberts1963,Rosenkilde1967} and gravitational
radiation~\citep{Chandrasekhar1970a}. These instabilities appear both
in Newtonian and general-relativistic setting and for realistic
equations of states, which indicates that they are generic in compact stars. The importance of the Chandrasekhar--Friedmann--Schutz (CFS)
instability~\citep{Chandrasekhar1970a,Friedman1978} to gravitational
wave-radiation in rotating stars, in particular, its manifestation in
the $r$-mode instability~(for reviews
see~\cite{Kokkotas2016,Andersson2021}), lies in the fact that
they may set an upper limit on rotational periods of a rapidly
rotating compact stars.

In a previous paper~\citep{Rau2020a} we showed that {\it
  Riemann ellipsoids} -- non-axisymmetric self-gravitating Newtonian
fluid bodies with constant internal circulation -- undergo secular
instabilities driven by gravitational wave
radiation or by viscosity (note, however, that viscosity alone may drive secular instability~\citep{Rosenkilde1967}).  Because these ellipsoids possess, in general, a non-vanishing fluid pattern in the frame rotating with its surface, they can be view as (approximate) models of differentially-rotating HMNS. Although more complex models
which include post-Newtonian corrections~\citep{Chandrasekhar1974,Shapiro1998,Gurlebeck2010,Gurlebeck2013}, 
full relativity or realistic equations of state can be
constructed, it is useful to first establish the key features in
the classical framework of ellipsoids, as exemplified by the cases 
 without internal circulation by~\cite{Chandrasekhar1969}.

 In this paper, we provide a detailed description of the formalism and
 extended computations within the approach adopted in~\cite{Rau2020a}. In
 particular, we focus on the unstable modes of Riemann S-type
 ellipsoids including gravitational radiation and shear viscosity. In
 doing so we find the modes for multiple stellar masses and various
 choices of shear viscosity.  While in~\cite{Rau2020a} we studied
 secularly unstable modes of $2.74M_{\odot}$, uniform density Riemann
 S-type ellipsoids, in this paper we cover a range of masses between
 $2M_{\odot}$, corresponding to a neutron star below the maximum mass
 at which collapse to a black hole is inevitable, up to
 $3.5M_{\odot}$, which is roughly at the mass limit for prompt
 collapse to a black hole. We also examine a range of (shear) viscosities, which includes ``enhanced'' values (aimed to mimic the effect of putative turbulent viscosity) which lead to dissipation comparable to the gravitational radiation damping. We assume that HMNS are sufficiently hot so that the superfluidity of baryonic matter which arises at low temperatures can be neglected; otherwise one needs to minimally account for the two-fluid nature of matter and for superfluid effects such as mutual friction, which can be included in the formalism adopted here [see~\cite{Sedrakian2001}].

 The non-dissipative modes of oscillations of Riemann ellipsoids were already derived in EFE. However, subsequent work on Riemann ellipsoids 
 focused on a different problem - the modeling of the secular evolution of differentially-rotating stars under the action of gravitational radiation and (shear)
 viscosity. Clearly, Riemann ellipsoids undergo unstable evolution of their triaxial shape
due to gravitational radiation.~\citet{Press1973} showed, through numerical
 integration of the equations of motion which included viscosity of
 the fluid, that secular instability drives a Maclaurin spheroid into
 a stable Jacobi ellipsoid via intermediate states which are Riemann
 S-type ellipsoids. The equations of motion of Riemann S-type
 ellipsoids under gravitational radiation-reaction were later
 integrated by \cite{Miller1974} and it was shown that the
 they again evolve into bodies with vanishing internal circulation
 with or without axial symmetry. The combined effect of viscosity and
 gravitational radiation was first considered by~\cite{Detweiler1977} and further extended by~\cite{Lai1995} to the compressible ellipsoidal
 approximation~\citep{Lai1993} to obtain insights into the
 evolution of a secularly unstable newly-born neutron star.
 The dissipative modes of Riemann ellipsoids were derived only recently
 \citep{Rau2020a}; here we provide an
 extended discussion of the underlying formalism and a parameter study,
 which complements our earlier discussion \citep{Rau2020a}. 

  Section~\ref{sec:TVFormalism} describes the tensor-virial formalism used to find the gravitational radiation-unstable modes and determine the effects of viscosity on them. Section~\ref{sec:NumericalResults} discusses the numerical results, by first 
  comparing the unstable mode growth times
  for different masses of models, and then for 
  a range of viscosity values. The
  results are summarized in Section~\ref{sec:Conclusion}. The
  gravitational wave back-reaction terms and characteristic equations
  used to compute the modes are given in full detail in
  Appendices~\ref{app:GWBackReaction1} and~\ref{app:CharEqs}
  respectively. In Appendix~\ref{app:GWBackReaction2} we provide the full 2.5-post-Newtonian gravitational radiation back-reaction terms for Riemann S-type ellipsoids, though as we discuss later on, only the limiting 
  expressions given in Appendix~\ref{app:GWBackReaction1} are used 
  for numerical computations. 

\section{Perturbation equations from the tensor-virial formalism}
\label{sec:TVFormalism}

The theory of the equilibrium ellipsoids and their oscillations 
is summarized by Chandrasekhar
in EFE.  We now briefly review the formalism used in the previous work 
~\citep{Rau2020a}, which is
based on EFE, and include below explicitly some of the equations which were
left out from this work for brevity.

We consider the perturbations of triaxial Riemann S-type ellipsoids
i.e. ellipsoids with principal axes $a_1\neq a_2\neq a_3$. The
principal axes are at rest in a corotating frame, which has angular
velocity $\boldsymbol{\Omega}=\boldsymbol{\Omega}(t)$ with respect to
the inertial frame, and which has internal motions with uniform
vorticity $\boldsymbol{\omega}$ as measured in the corotating
frame.  It is assumed that $\boldsymbol{\omega}$ and $\boldsymbol{\Omega}$ are parallel,
and are chosen to lie along the $z=x_3$ axis, which is the same in the
inertial and corotating frames. Without loss of generality we take
$a_1\geq a_2$. We consider incompressible flows
$\vecnabla\cdot\vecxi=0$, where $\vecxi$ is the Lagrangian displacement,
and assume uniform density for
simplicity. For perturbations with Lagrangian displacement of
the form
\begin{equation}
\vecxi(\mathbf{x},t)=e^{\lambda t}\vecxi(\mathbf{x}),
\label{eq:LagrangianPerturbation}
\end{equation}
the second-order tensor-virial equation leads to the characteristic equations
(including viscosity and
gravitational radiation back-reaction) given by (see also EFE)
\begin{align}
{}&\lambda^2V_{i;j}-2\lambda Q_{jl}V_{i;l}-2\lambda\Omega\epsilon_{i\ell 3}V_{\ell;j} \nonumber
\\
{}&-2\Omega\epsilon_{i\ell 3}(Q_{\ell k}V_{j;k}-Q_{jk}V_{\ell;k})+Q^2_{j\ell}V_{i;\ell}+Q^2_{i\ell}V_{j;\ell} \nonumber
\\={}&\Omega^2(V_{ij}-\delta_{i3}V_{3j})+\delta\mathfrak{W}_{ij}+\delta_{ij}\delta\Pi-\delta
\mathfrak{P}_{ij}-\delta\mathcal{G}_{ij},
\label{eq:CharacteristicEquation}
\end{align}
where the Latin indices $i,j =  1,2, 3$ are the components of the
Cartesian coordinate system.
This is a set of nine equations for $V_{i;j}$ and $V_{ij}$, 
which  are the unsymmetrized and symmetrized perturbations of the quadrupole moment tensor
\begin{equation}
V_{i;j}=\int_{\mathcal{V}} \text{d}^3x\rho\xi_ix_j \qquad V_{ij}=\int_{\mathcal{V}} \text{d}^3x\rho\left(\xi_ix_j+x_i\xi_j\right), \label{eq:VTensorDefinition}
\end{equation}
where $\rho$ is the density of the star, $x_i$ the coordinates in the
rotating frame and $\mathcal{V}$ the volume of the ellipsoid. The
matrices $Q_{ij}$ relate the background flow velocity inside the star
$u_i$ to the coordinates in the rotating frame $x_j$
\begin{equation}
u_i=Q_{ij}x_j,
\label{eq:BackgroundVelocity}
\end{equation}
where for the case of the Riemann S-type ellipsoids and with
$\boldsymbol{\Omega}$ and $\boldsymbol{\omega}$ aligned with the $x_3$
axis we have
\begin{align}
u_1={}&Q_{12}x_2, \qquad Q_{12}=-\frac{a_1^2}{a_1^2+a_2^2}\Omega f,
\label{eq:u1}
\\
u_2={}&Q_{21}x_1, \qquad Q_{21}=\frac{a_2^2}{a_1^2+a_2^2}\Omega f,
\label{eq:u2}
\\
u_3={}&0,
\label{eq:u3}
\end{align}
and all other elements of $Q_{ij}$ equal to zero.  The differential
rotation is parametrized in terms of the quantity
$f\equiv\omega/\Omega$, where $\omega$ and $\Omega$ are the magnitudes
of $\boldsymbol{\omega}$ and $\boldsymbol{\Omega}$. The different
Riemann sequences are labeled by their value of $f$, with $f=0$ being
the ({\it uniformly rotating)} Jacobi ellipsoids and $f=\pm\infty$ being the Dedekind ellipsoids
(for details see EFE). The case $f=-2$ corresponds to an irrotational ellipsoid
since the vorticity in the inertial frame is given by 
$\boldsymbol{\omega}_0=(2+f)\boldsymbol{\Omega}$. In Eq.~\eqref{eq:CharacteristicEquation}
$\delta\mathfrak{W}_{ij}$ denotes the gravitational potential energy
tensor given by
\begin{align}
\delta\mathfrak{W}_{ij}=-\int_{\mathcal{V}}\text{d}^3x\rho\xi_{\ell}\frac{\partial\mathfrak{B}_{ij}}{\partial
  x_{\ell}},
  \label{eq:W_ij}
\end{align}
where $\mathfrak{B}_{ij}$ is defined such that
\begin{equation}
\frac{\partial\mathfrak{B}_{ij}}{\partial x_{\ell}}=-\delta_{i\ell}\frac{\partial\mathfrak{B}}{\partial x_j}-\delta_{j\ell}\frac{\partial\mathfrak{B}}{\partial x_i}-3\mathfrak{B}_{ijl},
\end{equation}
and where
\begin{align}
\frac{\partial\mathfrak{B}}{\partial x_i}={}&-G\int_{\mathcal{V}}\text{d}^3x\rho(\mathbf{x}')\frac{x_i-x_i'}{|\mathbf{x}-\mathbf{x}'|^3},
\\
\mathfrak{B}_{ijl}={}&G\int_{\mathcal{V}}\text{d}^3x\rho(\mathbf{x}')\frac{(x_i-x_i')(x_j-x_j')(x_{\ell}-x_{\ell}')}{|\mathbf{x}-\mathbf{x}'|^5},
\end{align}
where $G$ is Newton's constant.
For a homogeneous star, EFE gives
\begin{equation}
\mathfrak{B}_{ij}=\pi G\rho\left[2B_{ij}x_ix_j+a_i^2\delta_{ij}\left(A_i-\sum_{\ell=1}^3A_{i\ell}x^2_{\ell}\right)\right],    
\label{eq:BijSymbol}
\end{equation}
where the index symbols $A_{ij}$ and $B_{ij}$ are defined as
\begin{align}
A_{ij}={}&a_1a_2a_3\int_0^{\infty}\frac{\text{d}u}{\Delta(u)(a_i^2+u)(a_j^2+u)},
\\
B_{ij}={}&a_1a_2a_3\int_0^{\infty}\frac{u\text{d}u}{\Delta(u)(a_i^2+u)(a_j^2+u)},
\\
\Delta(u)\equiv{}&\sqrt{(a_1^2+u)(a_2^2+u)(a_3^2+u)}.
\end{align}
It follows that in the homogeneous case Eq.~\eqref{eq:W_ij} gives 
\begin{align}
\delta\mathfrak{W}_{ij}=-\pi G\rho\Big[2B_{ij}V_{ij}-\delta_{ij}a_i^2\sum^3_{\ell=1}A_{i\ell}V_{\ell\ell}\Big].
\end{align}
Finally, 
$\delta\Pi$ In Eq.~\eqref{eq:CharacteristicEquation} is the Eulerian 
perturbation of the volume integral of the pressure, i.e., 
\begin{equation}
\delta\Pi=-\int_{\mathcal{V}}\text{d}^3x(\gamma-1)P\frac{\partial\xi_k}{\partial x_k},
\end{equation}
where $P=P(\mathbf{x})$ is the pressure and $\gamma$ is the ratio of specific heats. For incompressible flows, we must first eliminate $\delta\Pi$ from the virial characteristic equations and only then impose incompressibility.

The Eulerian perturbation of viscous stress tensor  $\delta\mathfrak{P}_{ij}$ for an incompressible fluid with a background velocity $u_k$ is
\begin{align}
\delta\mathfrak{P}_{ij}={}&\int_{\mathcal{V}}\text{d}^3x\rho\nu\Bigg[ \lambda\left(\frac{\partial\xi_i}{\partial x_j}+\frac{\partial\xi_j}{\partial x_i}\right)+\frac{\partial\xi_i}{\partial x_k}\frac{\partial u_k}{\partial x_j}
\nonumber
\\
{}&
+\frac{\partial\xi_j}{\partial x_k}\frac{\partial u_k}{\partial x_i}
+u_k\left(\frac{\partial^2\xi_i}{\partial x_k\partial x_j}+\frac{\partial^2\xi_i}{\partial x_k\partial x_j}\right)
\nonumber
\\
{}&
-\frac{\partial\xi_k}{\partial x_j}\frac{\partial u_i}{\partial x_k}-\frac{\partial\xi_k}{\partial x_i}\frac{\partial u_j}{\partial x_k}\Bigg]
\label{eq:ViscosityTerm}
\end{align}
where $\nu$ is the kinematic shear viscosity coefficient. We note
that in~\cite{Rau2020a} we kept only the the first term in the
integrand and thus neglected (erroneously) additional terms due to
internal rotation; we keep these terms here and verify that the
results reported in ~\cite{Rau2020a} are unchanged qualitatively. Below we
work in the low Reynolds number approximation i.e. the laminar flow regime for which the displacement fields for the inviscid flow are essentially unchanged when viscosity is included.  This implies that we can use as the eigenfunctions $\boldsymbol{\xi}$ in the absence
of viscosity 
\begin{equation}
\xi_i=\sum^3_{m=1}L_{i;m}x_m,   
\label{eq:Eigenfunctions}
\end{equation}
when evaluating the perturbations of the viscous stress tensor; 
here the $L_{i;m}$ are nine constants determined in the non-dissipative limit. Inserting this into Eq.~(\ref{eq:ViscosityTerm}) and using Eq.~(\ref{eq:BackgroundVelocity}) gives
\begin{align}
\delta\mathfrak{P}_{ij}=5\nu\Bigg[{}&\lambda\left(\frac{V_{i;j}}{a_j^2}+\frac{V_{j;i}}{a^2_i}\right)+Q_{kj}\frac{V_{i;k}}{a_k^2}-Q_{jk}\frac{V_{k;i}}{a_i^2}
\nonumber
\\
{}&+Q_{ki}\frac{V_{j;k}}{a_k^2}-Q_{ik}\frac{V_{k;j}}{a_j^2}\Bigg]. 
\end{align}
Note that Eq.~(\ref{eq:Eigenfunctions}) implies that for incompressible flows
\begin{equation}
\frac{V_{11}}{a_1^2}+\frac{V_{22}}{a_2^2}+\frac{V_{33}}{a_3^2}=0.
\label{eq:IncompressibilityCondition}
\end{equation}
According to Appendix~\ref{app:GWBackReaction1}, the gravitational
radiation back-reaction term $\delta\mathcal{G}_{ij}$ is
\begin{equation}
\delta\mathcal{G}_{ij}=\frac{2G}{5c^5}\left(\msout{I}^{(5)}_{i\ell}V_{\ell j}+\msout{V}^{(5)}_{i\ell}I_{\ell j}\right),
\label{eq:DeltaGijMainText}
\end{equation}
where $\msout{I}^{(5)}_{ij}$ is the fifth time derivative of the reduced quadrupole moment tensor of the ellipsoid in the inertial frame projected onto the rotating frame, defined in terms of the quadrupole moment tensor as
\begin{equation}
\msout{I}^{(5)}_{ij}=I^{(5)}_{ij}-\frac{1}{3}\delta_{ij}\text{Tr}(I^{(5)}),
\end{equation}
where~\citep{Chandrasekhar1970,Miller1974}
\begin{equation}
I^{(5)}_{ij}=\sum^{5}_{m=0}\sum^{m}_{p=0}C^{5}_{m}C^{m}_{p}(-1)^{p}[(\overline{\boldsymbol{\Omega}}^*)^p]_{ik}\frac{\text{d}^{5-m}I^{(r)}_{k\ell}}{\text{d}t^{5-m}}
[(\overline{\boldsymbol{\Omega}}^*)^{m-p}]_{\ell j}.
\label{eq:MomentofInertiaInertialFrame}
\end{equation}
$I^{(r)}_{k\ell}$ is the moment of inertia tensor in the rotating frame, $C^{m}_{n}$ are binomial coefficients, and for rotation about the $x_3$ axis, the matrix $\overline{\boldsymbol{\Omega}}^*$ takes the form
\begin{equation}
\overline{\boldsymbol{\Omega}}^*=\left(\begin{array}{ccc} 0 & \Omega & 0 \\ -\Omega & 0 & 0 \\ 0 & 0 & 0 \end{array} \right)\equiv\Omega\boldsymbol{\sigma},
\end{equation}
which defines the matrix $\boldsymbol{\sigma}$. For a time-independent moment of inertia as measured in the rotating frame, Eq.~(\ref{eq:MomentofInertiaInertialFrame}) reduces to
\begin{equation}
I^{(5)}_{ij}=\sum^{5}_{p=0}C^{5}_{p}(-1)^{p}[(\overline{\boldsymbol{\Omega}}^*)^p]_{ik}I^{(r)}_{k\ell}[(\overline{\boldsymbol{\Omega}}^*)^{5-p}]_{\ell j}.
\label{eq:MomentofInertiaInertialFrame2}
\end{equation}
For a triaxial ellipsoid in the rotating frame, where the principal
axes are aligned with the coordinate axes, $I_{ij}=I_{ij}^{(r)}$ is
\begin{equation}
I_{ij}=\frac{1}{5}M\delta_{ij}\left(\sum^3_{m=1}a_m^2-a_i^2\right).
\end{equation}
We further use the expression
for $\msout{I}^{(5)}_{ij}$  given in the appendix of
~\cite{Lai1994a}. Since $\Omega$ and $I_{ij}$
are constant in time, the only nonzero component of $\msout{I}^{(5)}_{ij}$ is
\begin{equation}
\msout{I}^{(5)}_{12}=\msout{I}^{(5)}_{21}=16\Omega^2(I_{11}-I_{22}).
\end{equation}
Analogously to $\msout{I}^{(5)}_{ij}$ one finds for $\msout{V}^{(5)}_{ij}$
\begin{equation}
\msout{V}^{(5)}_{ij}=V^{(5)}_{ij}-\frac{1}{3}\delta_{ij}\text{Tr}(V^{(5)}),
\end{equation}
where
\begin{align}
V^{(5)}_{ij}={}&\sum^{5}_{m=0}\sum^{m}_{p=0}C^{5}_{m}C^{m}_{p}(-1)^{p}[(\overline{\boldsymbol{\Omega}}^*)^p]_{ik}\frac{\text{d}^{5-m}V_{k\ell}}{\text{d}t^{5-m}}[(\overline{\boldsymbol{\Omega}}^*)^{m-p}]_{\ell j} \nonumber
\\
={}&\lambda^5V_{ij}-20\lambda\Omega^2(\lambda^2-2\Omega^2)(V_{ik}(\boldsymbol{\sigma}^4)_{kj}+\sigma_{ik}V_{k\ell}\sigma_{\ell j}) \nonumber
\\{}&+\Omega(5\lambda^4-40\lambda^2\Omega^2+16\Omega^4)(V_{ik}\sigma_{kj}-\sigma_{ik}V_{kj})
\nonumber
\\
={}&\lambda^5V_{ij}-\phi_1(\lambda,\Omega)(V_{ik}(\boldsymbol{\sigma}^4)_{kj}+\sigma_{ik}V_{k\ell}\sigma_{\ell j}) \nonumber
\\{}&+\phi_2(\lambda,\Omega)(V_{ik}\sigma_{kj}-\sigma_{ik}V_{kj}),
\label{eq:MomentofInertiaInertialFramePerturbation}
\end{align}
where we used the time-dependence of $\boldsymbol{\xi}$ from Eq.~(\ref{eq:LagrangianPerturbation}) and defined the auxiliary functions
\begin{align}
\phi_1(\lambda,\Omega)\equiv{}& 20\lambda\Omega^2(\lambda^2-2\Omega^2),
\\
\phi_2(\lambda,\Omega)\equiv{}& \Omega(5\lambda^4-40\lambda^2\Omega^2+16\Omega^4).
\end{align}
The $i,j\neq3$ components of $V^{(5)}_{ij}$ match those of
$\delta\mathbf{I}^{(5)}$
given in Eq.~(28) of \cite{Chandrasekhar1970}.

The rotational frequency and the characteristic frequencies can be made dimensionless using
\begin{equation}
\overline{\Omega}\equiv\frac{\Omega}{\sqrt{\pi G\rho}}, \qquad \overline{\lambda}\equiv\frac{\lambda}{\sqrt{\pi G\rho}}.
\label{eq:ReducedUnits}
\end{equation}
The shear viscosity term coefficients are made dimensionless through an additional factor of $a_1^{-2}$ such that the dimensionless kinematic shear viscosity is
\begin{align}
\overline{\nu}{}&\equiv\frac{\nu}{a_1^2\sqrt{\pi G\rho}}
\label{eq:ReducedShearViscosity}
\\
\nonumber
{}&=1.35\times10^{-13}\left(\frac{\nu}{10^3\text{ cm}^2\text{ s}^{-1}}\right)\left(\frac{\rho_0}{\rho}\right)^{1/2}\left(\frac{10\text{ km}}{a_1}\right)^2,
\end{align}
where $\rho_0=2.7\times10^{14}$ g/cm$^3$ is nuclear saturation density.
The terms describing the damping due to gravitational wave radiation will scale as $\overline{t}^{5}_c$, where $\overline{t}_c$ is the dimensionless light crossing time
\begin{equation}
\overline{t}_c\equiv\frac{a_1\sqrt{\pi G\rho}}{c}=0.251\left(\frac{\rho}{\rho_0}\right)^{1/2}\left(\frac{a_1}{10\text{ km}}\right).
\label{eq:ReducedLightCrossingTime}
\end{equation}
The dependence of $\overline{\nu}$ and $\overline{t}_c$ on the density
and the semi-axis length $a_1$ of the ellipsoid means that the mode
frequencies $\overline{\lambda}$ will also depend on these quantities,
unlike in the cases where viscosity and gravitational wave damping  
are absent.

\section{Numerical results}
\label{sec:NumericalResults}

\begin{figure}
    \centering
    \includegraphics[width=0.99\columnwidth]{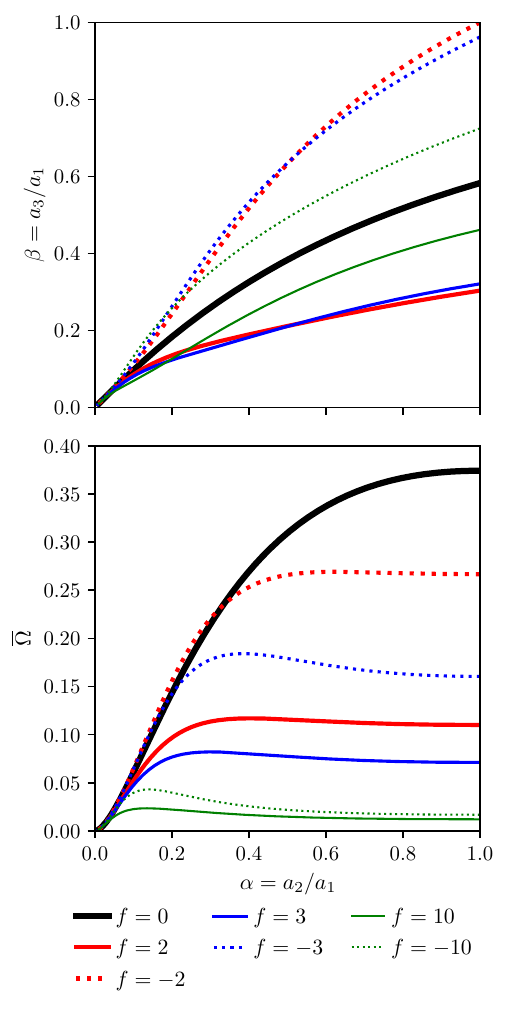}
    \caption{Equilibrium sequences of Riemann S-ellipsoids 
    parametrized by the reduced values of the semi-major axes $\alpha=a_2/a_1$ and $\beta=a_3/a_1$ for several values of circulation parameter $f$ (upper panel). 
The corresponding non-dimensional rotation frequency $\overline{\Omega}$
is shown in the lower panel. Note that $f=-2$ corresponds
     to the irrotational case and $f=0$ -- to the rigidly rotating case. 
     In the limit $f\rightarrow\pm\infty$ 
      the rotation frequency $\overline{\Omega}\rightarrow 0$.}
    \label{fig:EqSequences}
\end{figure}

As discussed in \cite{Rau2020a}, the procedure to compute the ellipsoid modes first involves determining the sequence of equilibrium Riemann S-type ellipsoids for each value of $f$ using the procedure described in EFE. These sequences consist of values of $\alpha$, $\beta$ and $\overline{\Omega}$. A selection of these sequences in the range $-\infty  \le f \le \infty$ are given in Fig.~\ref{fig:EqSequences}. The mode frequencies are then computed by solving matrix equations formed from the components of the tensorial characteristic equation Eq.~(\ref{eq:CharacteristicEquation}); these equations are given explicitly in Appendix~\ref{app:CharEqs}.

The components of the characteristic equation, and the resulting mode
frequencies, are grouped into even ($ij=11,22,33,12,21$) and odd
($ij=13,31,23,32$) in the index $3$ (the $x_3$ axis is aligned with the spin vector of the ellipsoid). The even modes correspond to
toroidal perturbations of the ellipsoid, and the odd modes to
transverse-shear perturbations. The even-parity equations result
in an order 17 polynomial in $\overline{\lambda}$, and the odd-parity equations in
a polynomial of order 14 in $\overline{\lambda}$. However, not all the 
modes are physically relevant. The reason is that, as in ~\cite{Rau2020a}, we work with Newtonian background ellipsoids and Newtonian equations of motion, therefore we are able to compute the gravitational wave back-reaction effects on the ``perturbative'' modes only. By this, we mean the modes which differ from the normal non-dissipative modes of the ellipsoid $\lambda_0$ by a small correction $\delta\lambda\in\mathbb{C}$, $|\delta\lambda|\ll|\lambda_0|$. Since 
the equilibrium background upon which 
perturbations are imposed is 
Newtonian, it is sufficient to consider the leading post-Newtonian
order gravitational-radiation reaction  
contribution given by Eq.~(\ref{eq:DeltaGijMainText}). 
However, in Appendix ~\ref{app:GWBackReaction2}, we compute the full 
2.5-post-Newtonian gravitational wave back-reaction terms. These can be use to address (some of) the remaining non-perturbative modes by computing the equilibrium background models at a post-Newtonian order. Such a program will allow one to assess the oscillation frequencies and damping of non-perturbative modes.

When discussing the numerical results for perturbative 
modes we define $\overline{\sigma}=-i\overline{\lambda}$. The unstable 
perturbative modes are those with $\text{Im}(\sigma)<0$, and the dimensionless growth 
time of these modes is specified by
\begin{equation}
    \overline{\tau}=-\frac{1}{\text{Im}(\overline{\sigma})}.
\end{equation}
 By definition, $\text{Re}(\sigma)$ provides the oscillation frequency of the mode.
We assign, additionally, indices $e$ or $o$ to quantities referring to even or odd modes respectively.

In our previous work \citep{Rau2020a} we included unstable modes for a 2.74$M_{\odot}$ ellipsoid (to match the mass of GW170817) with
uniform density $\rho=3.62\rho_0$. In this section we examine a range
of stellar masses $2M_{\odot}<M<3.5M_{\odot}$. \cite{Weih2018} gives a
threshold mass for prompt collapse as $1.54M_{\text{TOV}}$ for a
differentially-rotating star, so for a maximum TOV mass
$M_{\text{TOV}}$ in line with the highest measured neutron star mass
to date  of $2.14M_{\odot}$~\citep{Cromartie2020}, we should at least
consider masses above $\sim3.3M_{\odot}$, and thus choose an upper
mass range of 3.5$M_{\odot}$. For each model, we set the uniform
density $\rho$ by enforcing that the $f=-2$, $\alpha=\beta=1$ star has
a radius of $a_1=11$ km. For every other ellipsoid for a particular
fixed mass, we adjust $a_1$ so that each ellipsoid has constant
volume, and hence the same mass. The six stellar models examined in
the next two sections, and their uniform densities, are listed in
Table~\ref{tab:StellarModels}.

\begin{table}
    \centering
    \begin{tabular}{c|c}
         $M/M_{\odot}$ &  $\rho/\rho_0$\\
         \hline
         2 & 2.64 \\
         2.5 & 3.30 \\
          2.75 & 3.64 \\
          3 & 3.96 \\
          3.25 & 4.29 \\
           3.5 & 4.62 
    \end{tabular}
    \caption{The six stellar models studied, listing their masses and densities (in units of the nuclear saturation density $\rho_0=2.7\times10^{14}$ g/cm$^3$).}
    \label{tab:StellarModels}
\end{table}

\subsection{Variable mass}

\begin{figure}
    \centering
    \includegraphics[width=\columnwidth]{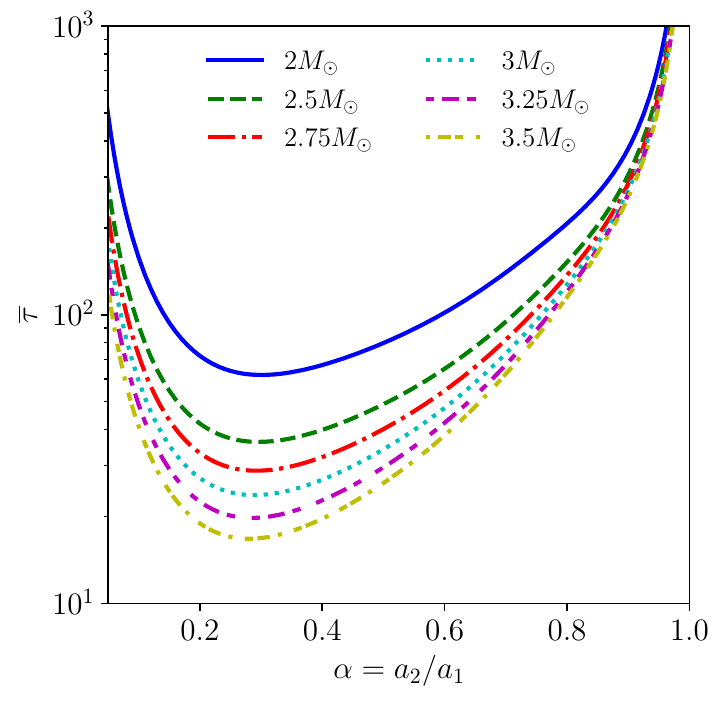}
    \includegraphics[width=\columnwidth]{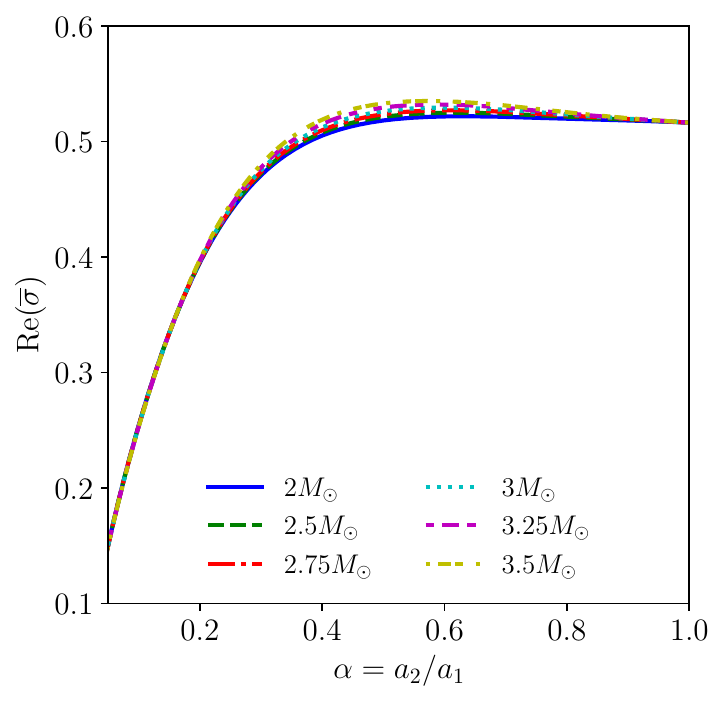}
    \caption{Growth time of unstable (odd) modes $\tau_o$ in reduced units for variable mass, $f=-2$, and $\nu=10^{14}$ cm$^2$/s (upper panel) and the corresponding oscillation frequencies $\text{Re}(\overline{\sigma}_o)$ (lower panel).}
    \label{fig:GrowthTimesf-2}
\end{figure}

\begin{figure}
    \centering
    \includegraphics[width=\columnwidth]{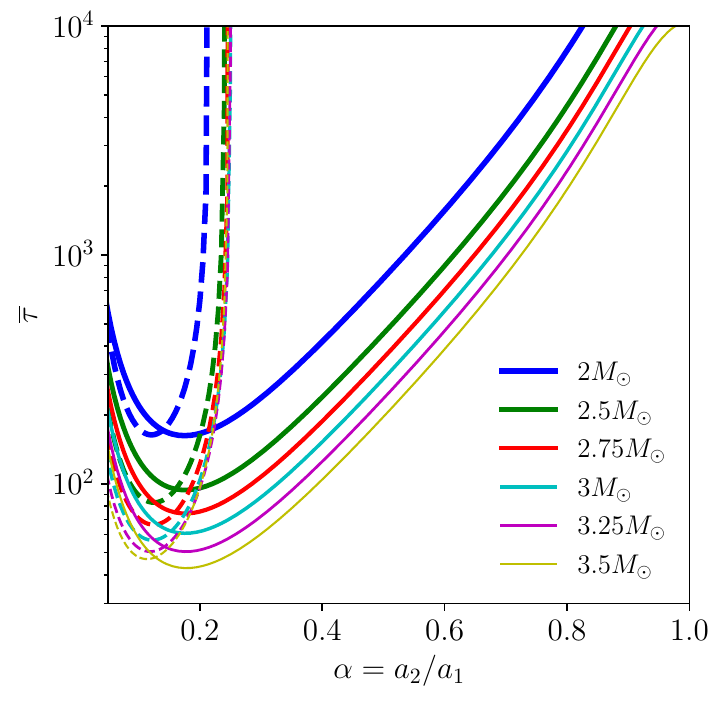}
    \includegraphics[width=\columnwidth]{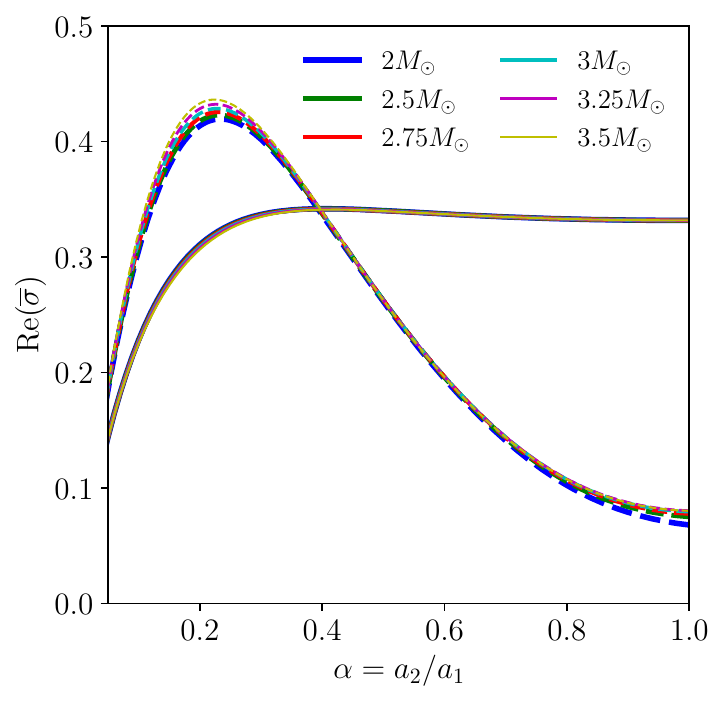}
    \caption{Same as in Fig.~\ref{fig:GrowthTimesf-2} except for $f=2$. 
    In each panel solid lines show the relevant quantities 
    ($\tau$ or $\text{Re}(\overline{\sigma})$)
    for odd modes and dashed lines - for even modes.
    }        \label{fig:GrowthTimesf2}
\end{figure}
The upper panel of Fig.~\ref{fig:GrowthTimesf-2}  shows the
growth times of the unstable modes for each of the chosen stellar-mass
models for $f=-2$ and with fixed viscosity
$\nu=10^{14}$ cm$^2$/s. The oscillation
frequencies corresponding to these unstable modes are shown in the lower panel of the same figure. Note the absence of
the instability for the even modes for $f=-2$, which is true generally
for $f<0$. Also note that the oscillation frequencies are almost independent of the mass, which is consistent with our restriction to studying the perturbative modes with $\text{Re}(\sigma_0)\gg|\delta\sigma|$, where $\sigma_0$ corresponds to the non-dissipative limit.  Figure~\ref{fig:GrowthTimesf2} is identical to the 
previous figure except now  $f=2$. The minimum growth times for the $f=2$ unstable modes occur at $\alpha\approx0.25$, and correspond to numerical values (restoring dimensionality) of $1.1$--$4.9$ ms. For $f=2$, the minimum growth times
for the unstable even and odd modes are similar, with the minimum
occurring at $\alpha\approx0.18$ for the unstable odd modes and at
$\alpha\approx0.11$ for the unstable even modes.  The unstable odd
modes are unstable for all $\alpha$, with the growth time increasing
as $\alpha\rightarrow1$. In the $f=-2$ case, $\tau_o$ approaches
infinity as $\alpha\rightarrow1$ since the $f=-2$ ellipsoid does not
have a mass quadrupole moment when $\alpha=\beta=1$. Changing the mass
has a very modest effect on the oscillation frequencies as expected
from our imposed restriction to the perturbative
modes with $\text{Re}(\sigma)\approx\text{Re}(\sigma_0)$, since the
modes of the undamped star (computed in EFE) are independent of the
stellar density and hence should only be slightly modified by the
gravitational radiation back-reaction.

\begin{figure}
    \centering
\includegraphics[width=\columnwidth]{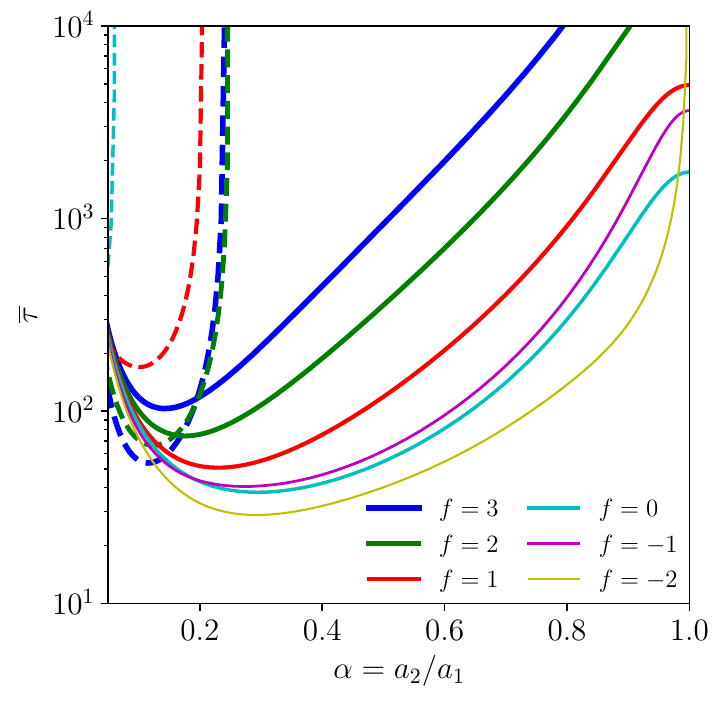}
\includegraphics[width=\columnwidth]{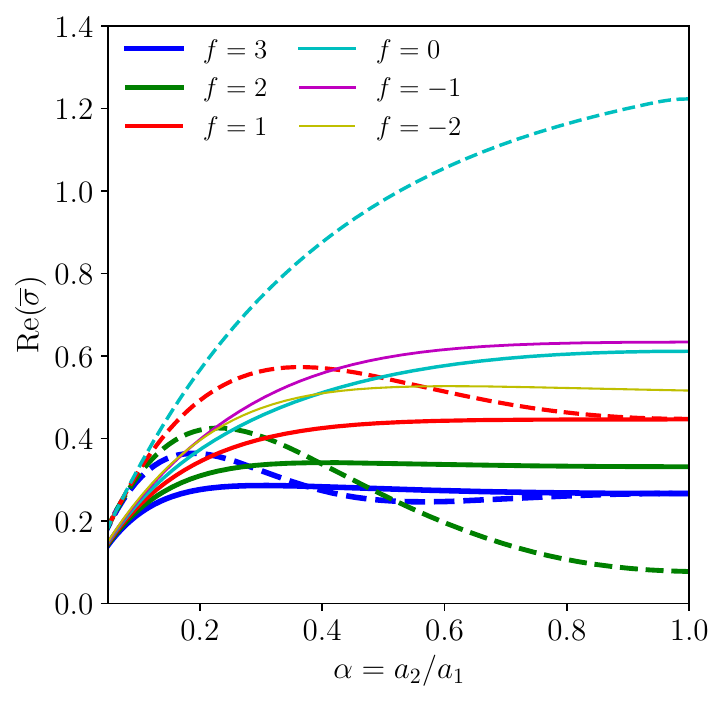}
    \caption{Growth time of unstable modes in reduced units for variable $f$ with $M=2.75M_{\odot}$ and $\nu=10^{14}$ cm$^2$/s (upper panel) and the 
    corresponding oscillation frequencies $\text{Re}(\overline{\sigma})$ (lower panel). Solid lines represent $\tau_o$ or $\text{Re}(\overline{\sigma}_o)$; dashed lines $\tau_e$ or $\text{Re}(\overline{\sigma}_e)$. }
    \label{fig:GrowthTimesM2.75}
\end{figure}

\begin{figure}
    \centering
    \includegraphics[width=\columnwidth]{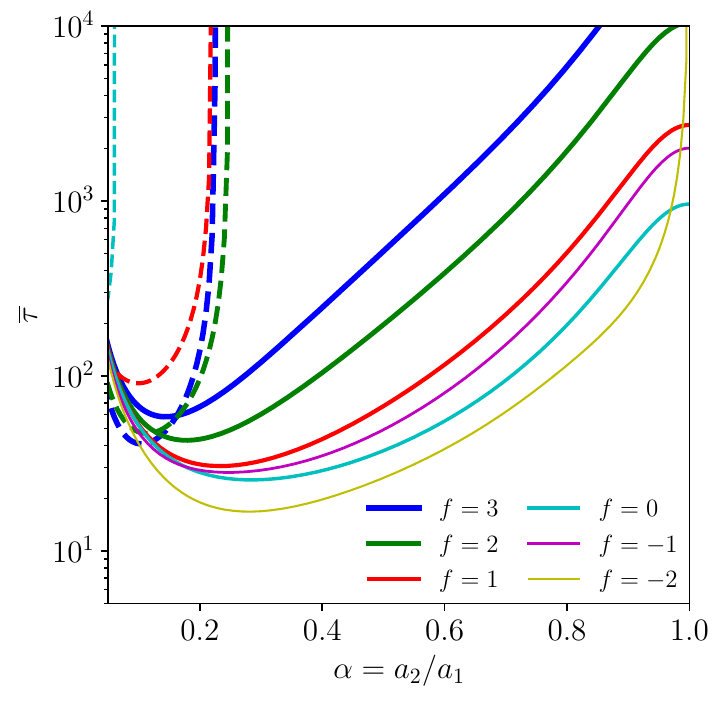}
    \includegraphics[width=\columnwidth]{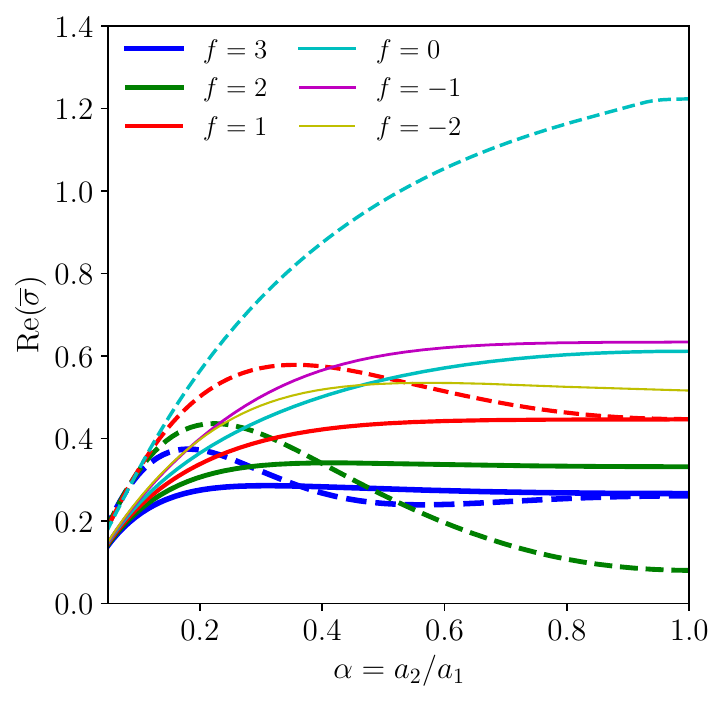}
    \caption{Same as Fig.~\ref{fig:GrowthTimesM2.75} except for $M=3.5M_{\odot}$.}
    \label{fig:GrowthTimesM3.5}
\end{figure}

Figures~\ref{fig:GrowthTimesM2.75}  and~\ref{fig:GrowthTimesM3.5} show in the upper panels the growth times of the unstable modes for the $M=2.75M_{\odot}$ and $M=3.5M_{\odot}$ stellar models for varying $f$. The even modes for $f<0$ are not unstable, though the instability in the $f=0$ case is of little physical interest since it only occurs for unreasonably eccentric ellipsoids. The instability of the even modes only occurs for highly eccentric ellipsoids $\alpha\lesssim0.25$, and their minimum growth times are shorter than the growth times for the unstable odd mode for the same stellar model for $f\gtrsim2$. The corresponding oscillation frequencies are shown in the lower panel of Figs.~\ref{fig:GrowthTimesM2.75}. For small $\alpha \simeq 0.1 $ the mode frequencies converge to the same value with distinct but similar values for odd and even modes. As $\alpha$ increases the odd modes saturate at a constant value starting from $\alpha \gtrsim 0.3$--$0.4$. The frequencies of the even modes pass through a maximum and decay for large $\alpha$ except for $f=0$ case which increases monotonically up to the point $\alpha =1$.

The GW-unstable odd modes occur for all equilibrium ellipsoids except the perfectly spherical $f=-2$, $\alpha=1$ model, unlike the even unstable modes which only appear for highly eccentric ellipsoids. When the viscosity is lowered, the instability for the less-eccentric ellipsoids can occur, as is shown in the next section. The presence or absence of this instability could hint at the sizes of the viscosities present in HMNS.

\subsection{Variable viscosity}

\begin{figure}
    \centering
    \includegraphics[width=\columnwidth]{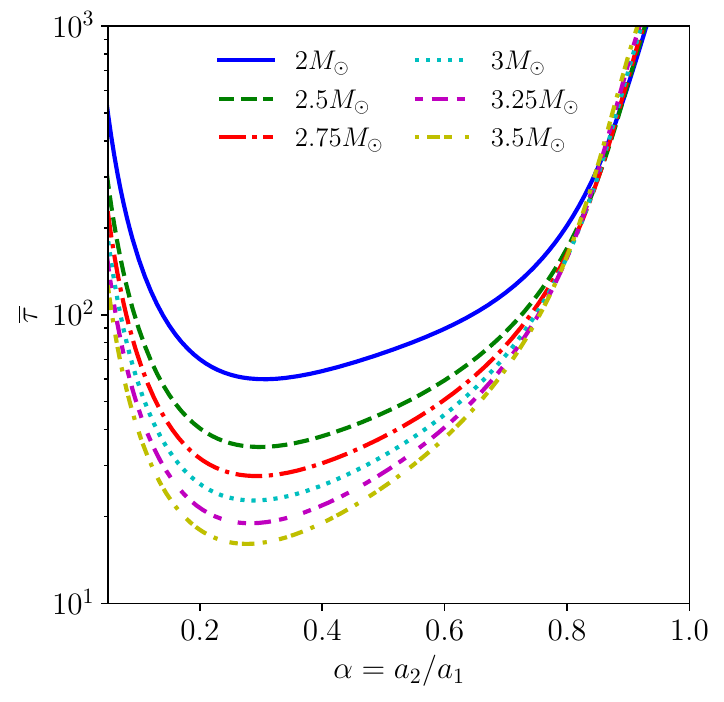}
     \includegraphics[width=\columnwidth]{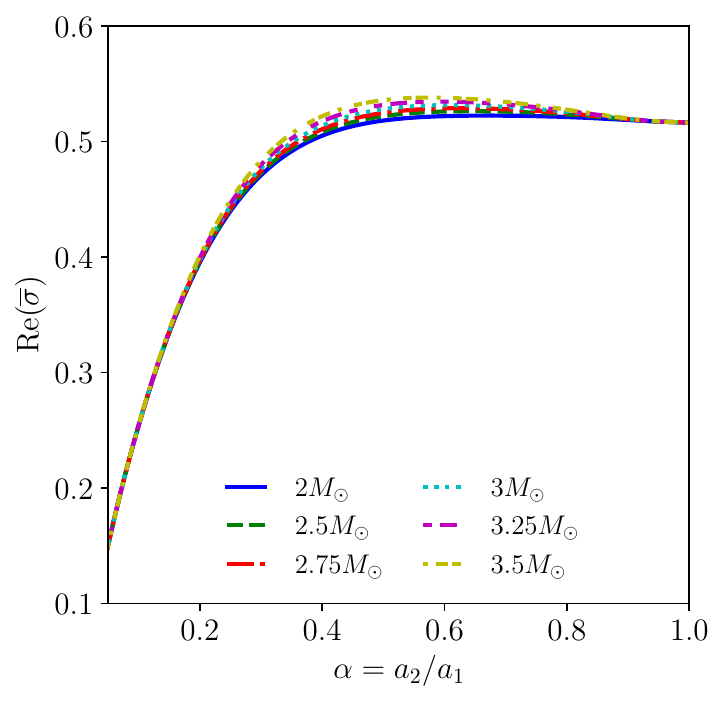}
    \caption{The growth times as in Fig.~\ref{fig:GrowthTimesf-2} except with $\nu=10^{13}$ cm$^2$~s$^{-1}$ (upper panel) and the corresponding oscillation frequencies $\text{Re}(\overline{\sigma})$ (lower panel). }
    \label{fig:GrowthTimesnu1e13f-2}
\end{figure}

\begin{figure}
    \centering
    \includegraphics[width=\columnwidth]{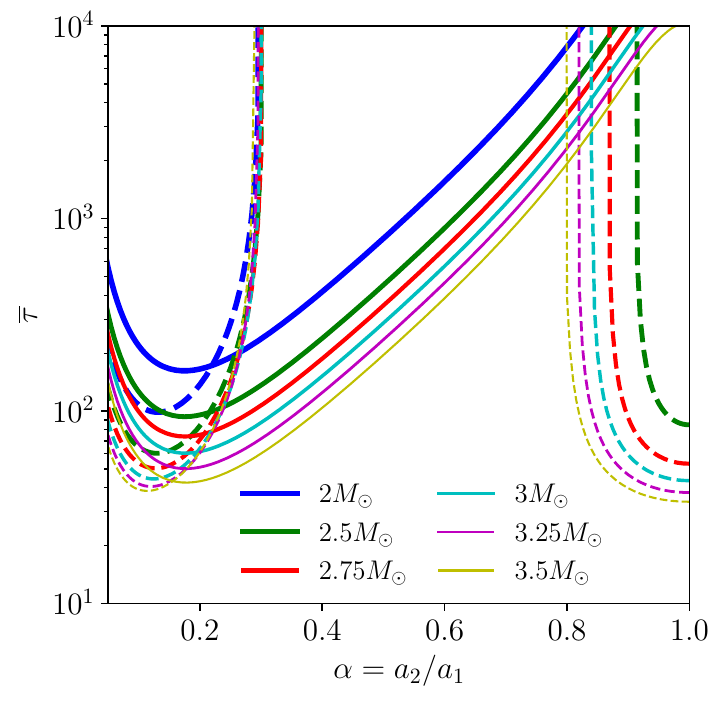}
    \includegraphics[width=\columnwidth]{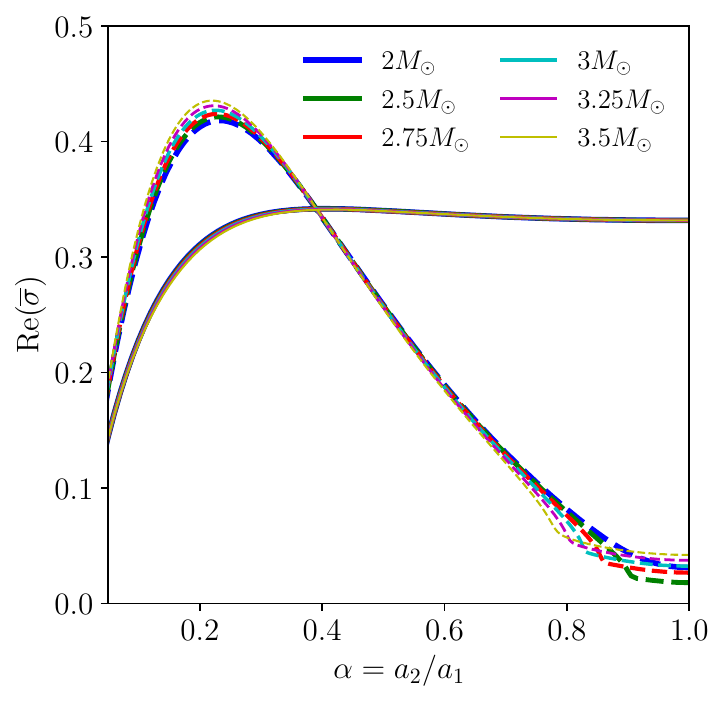}
    \caption{The growth times as in Fig.~\ref{fig:GrowthTimesf2} except with $\nu=10^{13}$ cm$^2$~s$^{-1}$. (upper panel) and the corresponding oscillation frequencies $\text{Re}(\overline{\sigma})$ (lower panel).}
    \label{fig:GrowthTimesnu1e13f2}
\end{figure}

As discussed in our previous work~\citep{Rau2020a}, exaggerated viscosities compared
to those that have been computed for neutron star interiors using
usual transport theory [for a review see~\citep{Schmitt2018}] 
are required for the viscosity to have any effect on the
modes. The required kinematic viscosities for physical relevance are
of order $10^{12}$--$10^{14}$cm$^2$~s$^{-1}$, which are far above the
typical values of $\nu\sim10$--$10^3$~cm$^2$s$^{-1}$ for a neutron
star core, assuming that the matter in the HMNS is similar to a
neutron star but at higher temperatures $T\sim10^{10}$--$10^{12}$
K. However, they are consistent with typical turbulent viscosities
used in astrophysical applications, including binary neutron star
merger simulations~\citep{Fujibayashi2018}. This is most often
implemented using the Shakura--Sunyaev $\alpha$-parameter
prescription~\citep{Shakura1973}
\begin{equation}
    \nu=\alpha_{\text{visc}} c_s H_{\text{turb}},
\end{equation}
where $\alpha_{\text{visc}}$ is the dimensionless $\alpha$-parameter, $c_s\sim c/3$ is the sound speed and $H_{\text{turb}}\sim 10^6$ cm is the turbulent eddy scale height, which should be of order the radius of the star. The range of nonzero viscosities we consider, $\nu=5\times10^{12}$--$10^{14}$ cm$^2$/s, are thus consistent with $\alpha_{\text{visc}}\sim0.0005$--$0.01$.

\begin{figure}
    \centering
    \includegraphics[width=\columnwidth]{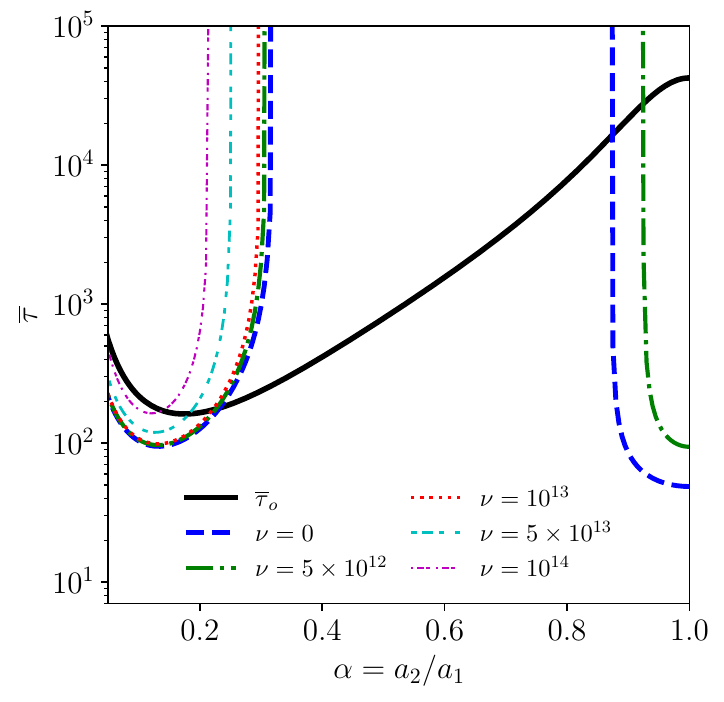}
    \includegraphics[width=\columnwidth]{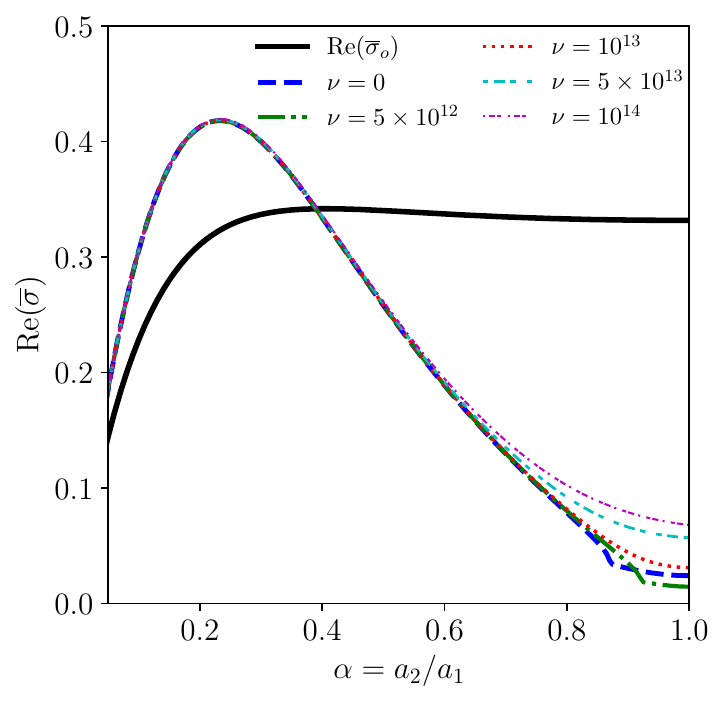}
    \caption{Growth time of unstable modes in reduced units for variable $\nu$ with $M=2M_{\odot}$ and $f=2$ (upper panel) and the corresponding oscillation 
    frequencies $\text{Re}(\overline{\sigma})$ (lower panel). The growth time of the unstable odd mode is not modified by changing the viscosity, and the $\tau_e$ and $\text{Re}(\sigma_e)$ are shown as non-solid lines. The viscosities are given in units of cm$^2$~s$^{-1}$.}
    \label{fig:GrowthTimesVarNuM2}
\end{figure}
\begin{figure}
    \centering
    \includegraphics[width=\columnwidth]{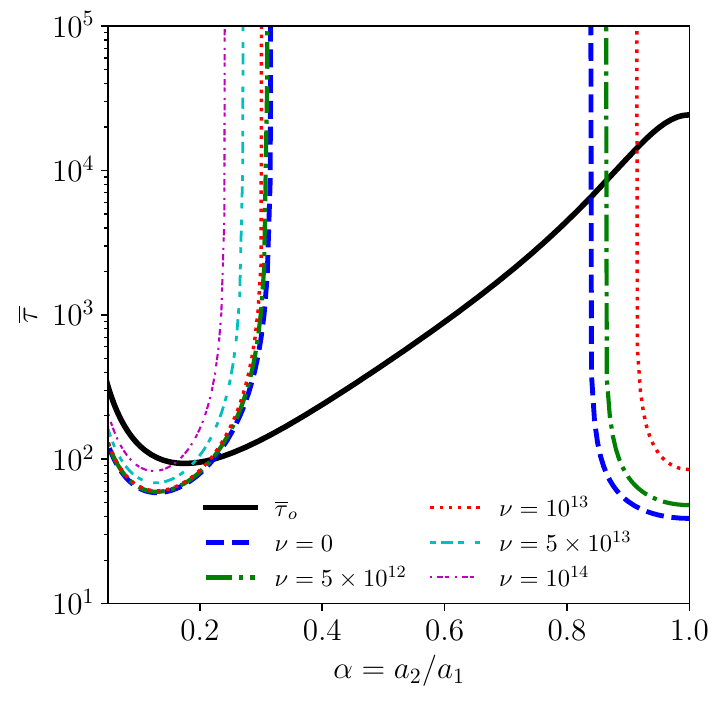}
    \includegraphics[width=\columnwidth]{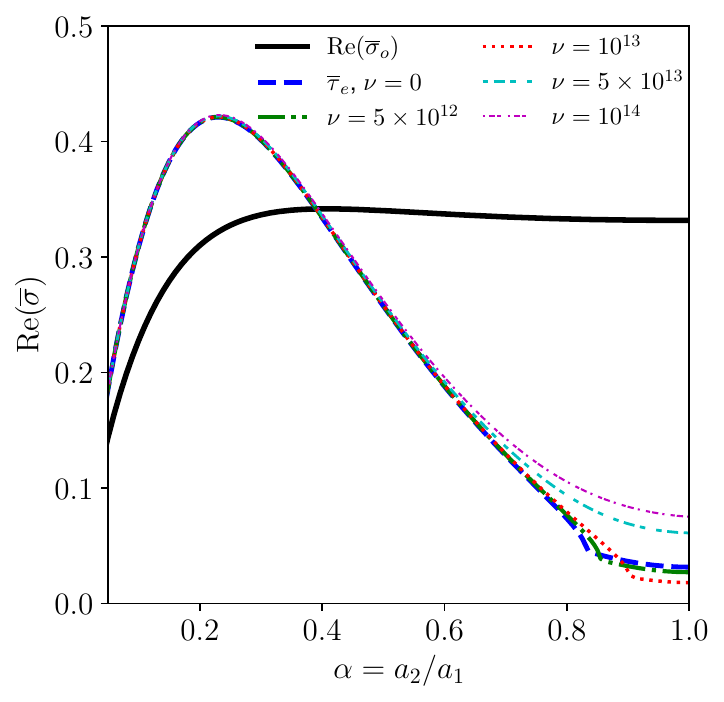}
    \caption{Same as Fig.~\ref{fig:GrowthTimesVarNuM2} except with $M=2.5M_{\odot}$.}
    \label{fig:GrowthTimesVarNuM2.5}
\end{figure}

Figures~\ref{fig:GrowthTimesnu1e13f-2} and~\ref{fig:GrowthTimesnu1e13f2} show in the upper panels the growth times of the unstable modes for each of the chosen stellar-mass models for $f=-2$ and $f=2$ respectively, with fixed viscosity $\nu=10^{13}$ cm$^2$/s. The corresponding oscillation frequencies are shown in the lower panels, respectively. Comparing to Figs.~\ref{fig:GrowthTimesf-2} and~\ref{fig:GrowthTimesf2}, we see that the growth times for the more eccentric ellipsoids $\alpha\lesssim0.4$ are nearly unaffected by the change in viscosity. This is to be expected since the quadrupole moment of the ellipsoid increases with eccentricity and the gravitational radiation thus dominates the viscosity. Most notably, lowering the viscosity opens up a second ``branch'' of instability for the unstable even modes for $f>0$: they are also unstable at large $\alpha\gtrsim0.8$ in addition to at $\alpha\lesssim0.25$, with comparable minimum growth times for both branches. However, at $\nu=10^{13}$ cm$^2$/s, the instability for $\alpha\gtrsim0.8$ is suppressed in the $2M_{\odot}$ ellipsoid: lowering the viscosity further allows this ellipsoid to be unstable in this range of $\alpha$ as shown later.

\begin{figure}
    \centering
    \includegraphics[width=\columnwidth]{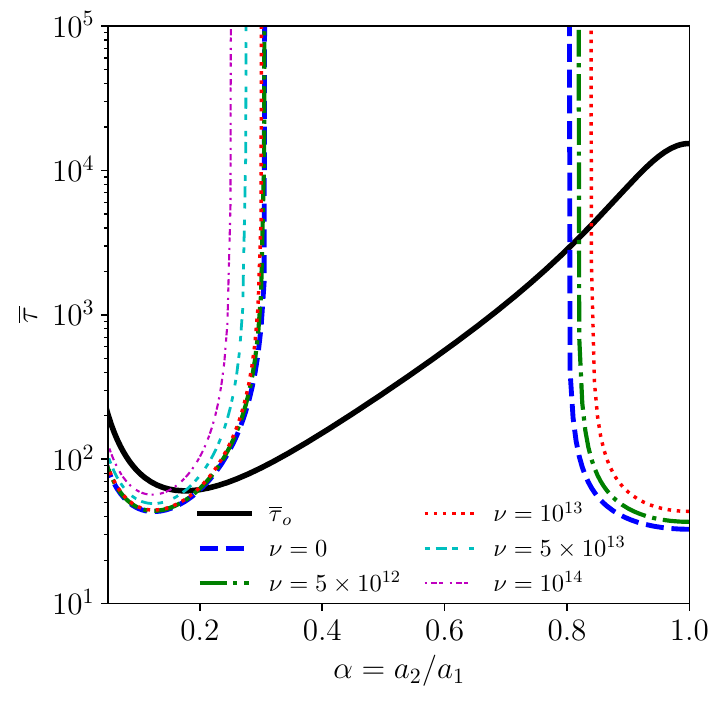}
     \includegraphics[width=\columnwidth]{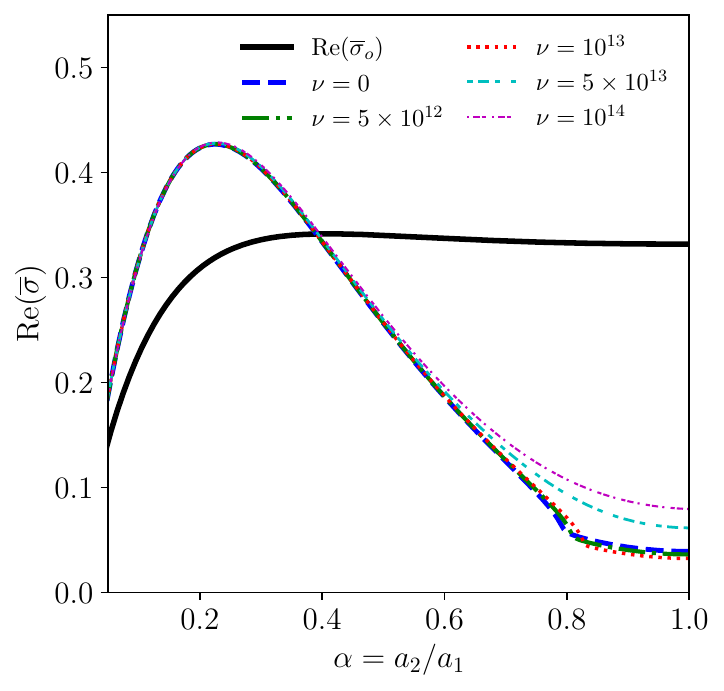}
    \caption{Same as Fig.~\ref{fig:GrowthTimesVarNuM2} except with $M=3M_{\odot}$.}
    \label{fig:GrowthTimesVarNuM3}
\end{figure}

\begin{figure}
    \centering
    \includegraphics[width=\columnwidth]{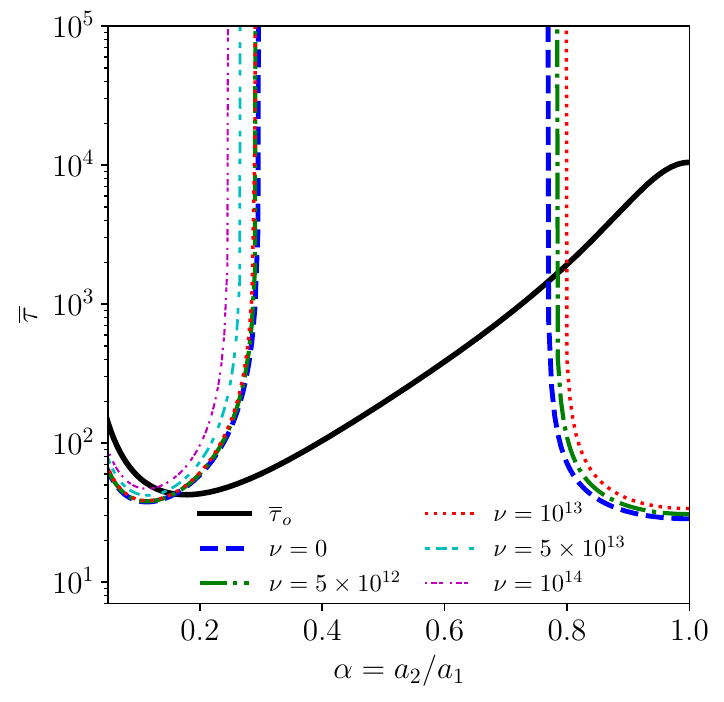}
    \includegraphics[width=\columnwidth]{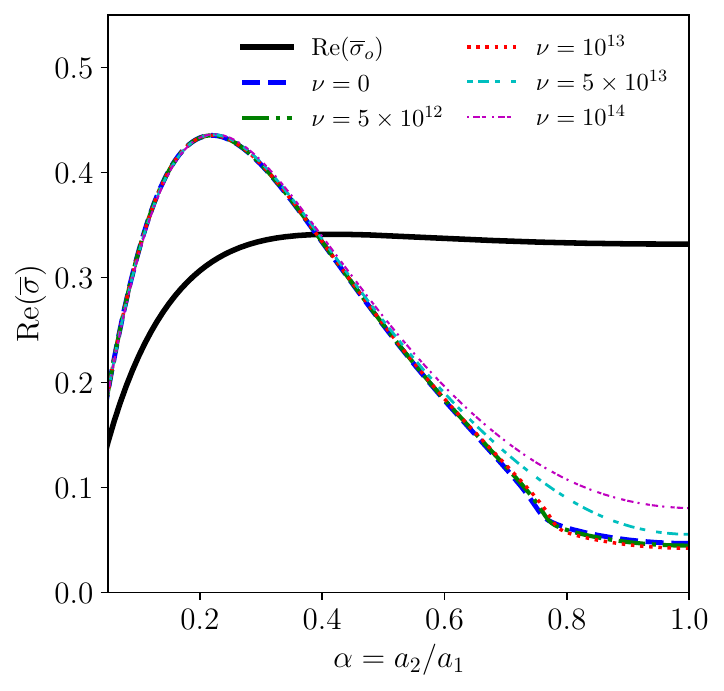}
    \caption{Same as FIG.~\ref{fig:GrowthTimesVarNuM2} except with $M=3.5M_{\odot}$.}
    \label{fig:GrowthTimesVarNuM3.5}
\end{figure}

Figure~\ref{fig:GrowthTimesVarNuM2}--\ref{fig:GrowthTimesVarNuM3.5} show the growth times and oscillation frequencies of the $f=2$ unstable modes for varying $\nu$ and  $M=2M_{\odot},2.5M_{\odot},3M_{\odot}$ and $3.5M_{\odot}$ stellar models respectively. The gravitational-radiation-unstable odd modes are unchanged by varying the viscosity for fixed $f$ and stellar mass. This is why only a single value of $\tau_o$ is included in each figure. In general, increasing the viscosity increases the growth times i.e. has a stabilizing effect as expected. Note that for the $M=2.5M_{\odot}$, $3M_{\odot}$ and $3.5M_{\odot}$ models, the high $\alpha$ branch of the instability vanishes above $\nu=10^{13}$ cm$^2$/s, while for the $M=2M_{\odot}$ model it vanishes above $\nu=5\times10^{12}$ cm$^2$/s. For $\alpha\lesssim0.75$, the values of $\text{Re}(\sigma_e)$ for the different choices of $\nu$  are equal to within a few percent, as expected for analogous reasons as to why $\text{Re}(\sigma)$ is only slightly changed as a function of mass for fixed $f$. For larger values of $\alpha$ there can be a significant difference in $\text{Re}(\sigma_e)$ as a function of viscosity

\section{Conclusion}
\label{sec:Conclusion}

In this paper, we have provided further details on the 
gravitational-radiation-unstable modes of Riemann S-type ellipsoids as computed
using the tensor-virial method, and made comparisons of the unstable
modes for different masses and viscosities. The range of masses
examined are appropriate for simple models of transient,
rapidly-rotating hypermassive neutron stars formed in BNS mergers. The
calculations here and in \cite{Rau2020a} give qualitative
estimates for the growth times and oscillation frequencies of the
unstable modes of HMNS, assuming they can settle into
quasi-stationary gravitational equilibria shortly after their birth
and before the collapse to a black hole.

As expected, the growth times of the unstable modes generally increase
as a function of the stellar mass. They are of order milliseconds, i.e., 
short enough to be relevant to HMNS. The growth times are increased by
viscosity, and its magnitude should be of order $\nu\simeq 10^{12}$ cm$^2$/s
or larger to have an observable effect on the unstable growth times for
the range of masses we considered. These values are impossible with
only standard viscosities computed using transport theory (e.g. using
the Chapman--Enskog expansion), but are attainable via turbulent
viscosity as applied in accretion disk theory and in numerical BNS
merger simulations. For $\nu\gtrsim10^{15}$ cm$^2$/s, the instability
can be suppressed completely. In general, the even unstable modes,
corresponding to toroidal perturbations, have slightly longer growth times for a given $f$,
mass and viscosity than the transverse-shear perturbation odd
modes. The even modes are unstable for highly eccentric ellipsoids
$\alpha=a_3/a_1\lesssim0.25$ for $f\geq0$, and can be unstable for
$\alpha\gtrsim0.8$ and $f\geq0$ if the viscosity is sufficiently
small. The odd modes are unstable for all $\alpha$ except the
spherical $\alpha=1$, $f=-2$ stellar model, and have growth times that
are minimized near $\alpha\approx0.25$.  

The insights gained here from the semi-analytical tensor-virial approach are expected to be useful when addressing the problem of HMNS oscillations in different settings and approximations, in particular when including such features as realistic equations of state, general relativity, and varying velocity profiles (with slowly rotating core and rapidly rotating envelope)
as seen in the numerical simulations of BNS mergers. We anticipate that the instabilities of the oscillation modes revealed in our analysis will be present in more realistic models, as has been the case for self-gravitating fluids without internal circulations. 

It is worthwhile to note that the oscillations of the type discussed here can be tested in the laboratory using ultracold atoms, for which the magnetic or laser trapping potential takes the role of the gravitational potential. For such systems, the tensor-virial method leads to a good agreement between the theory and experiment, as has been demonstrated in the case of the breathing modes of uniformly rotating clouds~\citep{Sedrakian2001,Watanabe2007PhRvA}.

\section{Acknowledgements}
We are grateful to I. Wasserman for discussions. AS acknowledges the support by the Deutsche Forschungsgemeinschaft (Grant No. SE 1836/5-1) and the European COST Action CA16214 PHAROS “The multi-messenger physics and astrophysics of neutron stars”.

\section*{Data Availability}

The data underlying this article will be shared on reasonable request to the corresponding author.

\bibliography{library,textbooks,librarySpecialPlusExtras}

\appendix

\section{Gravitational radiation back-reaction term in virial equation 1: Lowest order form}
\label{app:GWBackReaction1}

The gravitational radiation back-reaction potential $\Phi_{\text{GW}}$ in the weak-field, slow motion regime is~\citep{MTW1973,Lai1994a}
\begin{equation}
\Phi_{\text{react}}=-\frac{G}{5c^5}\msout{I}^{(5)}_{ij}x_ix_j.
\end{equation}
The corresponding gravitational radiation back-reaction force is given by~\citep{Miller1974}
\begin{align}
f_i^{\text{react}}=-\rho\frac{\partial\Phi_{\text{react}}}{\partial x^k}={}&\frac{2\rho G}{5c^2}\left(I^{(5)}_{ik}x_k-\frac{1}{3}I^{(5)}_{kk}x_i\right)
\nonumber
\\
={}&\frac{2\rho G}{5c^2}\msout{I}^{(5)}_{ik}x_k,
\end{align}
which leads to the second moment of this expression 
\begin{equation}
\mathcal{G}_{ij}=\frac{2G}{5c^5}\int_{\mathcal{V}}\text{d}^3x\rho\msout{I}^{(5)}_{ik}x_kx_j.
\end{equation}
Taking the perturbation of this tensor we find 
\begin{align}
\delta\mathcal{G}_{ij}={}&\frac{2G}{5c^5}\left(\msout{I}^{(5)}_{ik}V_{kj}+\int_{\mathcal{V}}\text{d}^3x\rho x_kx_j\delta\msout{I}^{(5)}_{ik}\right)
\nonumber
\\
={}&\frac{2G}{5c^5}\left(\msout{I}^{(5)}_{ik}V_{kj}+\msout{V}^{(5)}_{ik}I_{kj}\right),
\label{eq:DeltaGij}
\end{align}
where 
\begin{equation}
\delta I_{ij}=\delta \int_{\mathcal{V}}\text{d}^3x\rho x_ix_j=V_{ij}.
\end{equation}
and, since $\delta I_{ij}=\delta I^{(r)}_{ij}$ in a rotating frame,
\begin{align}
\delta I^{(5)}_{ij}={}&\sum^{5}_{m=0}\sum^{m}_{p=0}C^{5}_{m}C^{m}_{p}(-1)^{p}
\nonumber
\\
{}&\qquad\qquad\times[(\overline{\boldsymbol{\Omega}}^*)^p]_{ik}\frac{\text{d}^{5-m}\delta I^{(r)}_{k\ell}}{\text{d}t^{5-m}}
[(\overline{\boldsymbol{\Omega}}^*)^{m-p}]_{\ell j}
\nonumber
\\
\equiv{}& V^{(5)}_{ij}.
\end{align}
Eq.~(\ref{eq:DeltaGij}) includes the secular effects of gravitational radiation back-reaction for quadrupole radiation~\citep{Thorne1969}. Additional effects at higher orders in a post-Newtonian expansion can be incorporated using the formalism of~\citet{Chandrasekhar1970}. As we do not include the post-Newtonian effects on the background ellipsoid and in the tensor-virial formalism, it would be inconsistent to use these more-complicated gravitational radiation back-reaction terms in this paper.

\begin{onecolumn}
\section{Characteristic equations}
\label{app:CharEqs}

We explicitly write the nine different components of Eq.~(\ref{eq:CharacteristicEquation}) in terms of the $V_{ij}$ and $V_{i;j}$, the rotation frequency $\Omega$ and ratio $f$, the eigenvalue $\lambda$, the index symbols $B_{ij}$ and $A_{ij}$, and the mass and principal axes of the equilibrium ellipsoids. The five components even in the index 3 are

\begin{align}
\frac{1}{2}\lambda^2 V_{33}=
-\pi G\rho\left[ 2B_{33}V_{33}-a_3^2\sum^3_{\ell=1}A_{3\ell}V_{\ell\ell} \right]-\frac{5\lambda\eta}{a_3^2}V_{33}-\frac{2GM(a_1^2+a_2^2)}{75c^5}\lambda^5(2V_{33}-V_{11}-V_{22})+\delta\Pi,
\label{eq:CharacteristicEqn33}
\\
\left[\frac{1}{2}\lambda^2+Q_{12}Q_{21}-\Omega^2\right]V_{11}-2\lambda Q_{12}V_{1;2}-2\lambda\Omega V_{2;1}-\Omega(Q_{21}V_{11}-Q_{12}V_{22})=-\pi G\rho\left[2B_{11}V_{11}-a_1^2\sum^3_{\ell=1}A_{1\ell}V_{\ell\ell}\right]+\delta\Pi
\nonumber
\\-5\eta\left[\frac{\lambda V_{11}}{a_1^2}+2Q_{21}\frac{V_{1;2}}{a_2^2}-2Q_{12}\frac{V_{2;1}}{a_1^2}\right]
\nonumber
\\-\frac{2GM(a_2^2+a_3^2)}{25c^5}\left[\frac{1}{3}\lambda^5(2V_{11}-V_{22}-V_{33})-\phi_1(\lambda,\Omega)(V_{11}-V_{22})+\left(16\Omega^5\frac{a_2^2-a_1^2}{a_2^2+a_3^2}-2\phi_2(\lambda,\Omega)\right)V_{12}\right],
\label{eq:CharacteristicEqn11}
\\
\left[\frac{1}{2}\lambda^2+Q_{12}Q_{21}-\Omega^2\right]V_{22}-2\lambda Q_{21}V_{2;1}+2\lambda\Omega V_{1;2}+\Omega (Q_{12}V_{22}-Q_{21}V_{11})=-\pi G\rho\left[2B_{22}V_{22}-a_2^2\sum^3_{\ell=1}A_{2\ell}V_{\ell\ell}\right]+\delta\Pi
\nonumber
\\-5\eta\left[\frac{\lambda V_{22}}{a_2^2}+2Q_{12}\frac{V_{2;1}}{a_1^2}-2Q_{21}\frac{V_{1;2}}{a_2^2}\right]
\nonumber
\\-\frac{2GM(a_1^2+a_3^2)}{25c^5}\left[\frac{1}{3}\lambda^5(2V_{22}-V_{11}-V_{33})-\phi_1(\lambda,\Omega)(V_{22}-V_{11})+\left(16\Omega^5\frac{a_2^2-a_1^2}{a_1^2+a_3^2}+2\phi_2(\lambda,\Omega)\right)V_{12}\right],
\label{eq:CharacteristicEqn22}
\\
\lambda^2V_{1;2}-\lambda Q_{21}V_{11}-\lambda\Omega V_{22}=-5\eta\left[\lambda\left(\frac{V_{1;2}}{a_2^2}+\frac{V_{2;1}}{a_1^2}\right)+\frac{Q_{12}-Q_{21}}{2}\left(\frac{V_{11}}{a_1^2}-\frac{V_{22}}{a_2^2}\right)\right]
\nonumber
\\-\frac{2GM(a_1^2+a_3^2)}{25c^5}\left[16\Omega^5\frac{a_2^2-a_1^2}{a_1^2+a_3^2}V_{22}+(\lambda^5-2\phi_1(\lambda,\Omega))V_{12}+\phi_2(\lambda,\Omega)(V_{11}-V_{22})\right],
\label{eq:CharacteristicEqn12}
\\
\lambda^2V_{2;1}-\lambda Q_{12}V_{22}+\lambda\Omega V_{11}=-5\eta\left[\lambda\left(\frac{V_{1;2}}{a_2^2}+\frac{V_{2;1}}{a_1^2}\right)+\frac{Q_{12}-Q_{21}}{2}\left(\frac{V_{11}}{a_1^2}-\frac{V_{22}}{a_2^2}\right)\right]
\nonumber
\\-\frac{2GM(a_2^2+a_3^2)}{25c^5}\left[16\Omega^5\frac{a_2^2-a_1^2}{a_2^2+a_3^2}V_{11}+(\lambda^5-2\phi_1(\lambda,\Omega))V_{12}+\phi_2(\lambda,\Omega)(V_{11}-V_{22})\right],
\label{eq:CharacteristicEqn21}
\end{align}
where in the last two equations we used the relation $\Omega^2-Q_{12}Q_{21}=2B_{12}$ valid for Riemann ellipsoids with parallel $\omega$ and $\Omega$. The four components odd in the index 3 are
\begin{align}
\lambda^2 V_{1;3}-2\lambda\Omega V_{2;3}+Q_{12}Q_{21}V_{3;1}-2\Omega Q_{21}V_{3;1}={}&
(\Omega^2-2\pi G\rho B_{13})V_{13}-5\lambda\eta\left(\frac{V_{1;3}}{a_3^2}+\frac{V_{3;1}}{a_1^2}\right)
\nonumber
\\
{}&-\frac{2GM(a_1^2+a_2^2)}{25c^5}\lambda^5V_{13}-\frac{32GM(a_2^2-a_1^2)}{25c^5}\Omega^5V_{23},
\label{eq:CharacteristicEqn13}
\\
\lambda^2 V_{2;3}+2\lambda\Omega V_{1;3}+Q_{12}Q_{21}V_{3;2}+2\Omega Q_{12}V_{3;2}={}&
(\Omega^2-2\pi G\rho B_{23})V_{23}-5\eta\left(\lambda\frac{V_{2;3}}{a_3^2}+\lambda\frac{V_{3;2}}{a_2^2}+Q_{12}\frac{V_{3;1}}{a_1^2}-Q_{21}\frac{V_{1;3}}{a_3^2}\right)
\nonumber
\\
{}&-\frac{2GM(a_1^2+a_2^2)}{25c^5}\lambda^5V_{23}-\frac{32GM(a_2^2-a_1^2)}{25c^5}\Omega^5V_{13},
\label{eq:CharacteristicEqn23}
\\
\lambda^2 V_{3;1}-2\lambda Q_{12}V_{3;2}+Q_{12}Q_{21}V_{3;1}={}&
-2\pi G\rho B_{13}V_{13}-5\lambda\eta\left(\frac{V_{3;1}}{a_1^2}+\frac{V_{1;3}}{a_3^2}\right)
\nonumber
\\
{}&-\frac{2GM(a_2^2+a_3^2)}{25c^5}\left([\lambda^5-\phi_1(\lambda,\Omega)]V_{13}-\phi_2(\lambda,\Omega)V_{23}\right),
\label{eq:CharacteristicEqn31}
\\
\lambda^2 V_{3;2}-2\lambda Q_{21}V_{3;1}+Q_{12}Q_{21}V_{3;2}={}&
-2\pi G\rho B_{23}V_{23}-5\eta\left(\lambda\frac{V_{3;2}}{a_2^2}+\lambda\frac{V_{2;3}}{a_3^2}+Q_{12}\frac{V_{3;1}}{a_1^2}-Q_{21}\frac{V_{1;3}}{a_3^2}\right)
\nonumber
\\
{}&-\frac{2GM(a_1^2+a_3^2)}{25c^5}\left([\lambda^5-\phi_1(\lambda,\Omega)]V_{23}+\phi_2(\lambda,\Omega)V_{13}\right).
\label{eq:CharacteristicEqn32}
\end{align}
Eqs.~\eqref{eq:CharacteristicEqn33}--\eqref{eq:CharacteristicEqn32} can be divided by $\pi G\rho$, which makes  all the coefficients of $V_{i;j}$ and $V_{ij}$ dimensionless. The reduced equations are
\begin{align}
\frac{1}{2}\overline{\lambda}^2 V_{33}=
-\left[ 2\overline{B}_{33}V_{33}-\beta^2\sum^3_{\ell=1}\overline{A}_{3\ell}V_{\ell\ell} \right]-\frac{5\overline{\lambda}\overline{\eta}}{\beta^2}V_{33}-\frac{8\alpha\beta(1+\alpha^2)}{225}\overline{t}_c^5\overline{\lambda}^5(2V_{33}-V_{11}-V_{22})+\frac{\delta\Pi}{\pi G\rho},
\label{eq:CharacteristicEqn33R}
\\
\left[\frac{1}{2}\overline{\lambda}^2-\frac{\alpha^2\overline{\Omega}^2f^2}{(1+\alpha^2)^2}-\overline{\Omega}^2\right]V_{11}+\frac{2\overline{\Omega}\overline{\lambda}f}{1+\alpha^2}V_{1;2}-2\overline{\lambda}\overline{\Omega} V_{2;1}-\frac{\overline{\Omega}^2f}{1+\alpha^2}(\alpha^2V_{11}+V_{22})=-\left[2\overline{B}_{11}V_{11}-\sum^3_{\ell=1}\overline{A}_{1\ell}V_{\ell\ell}\right]+\frac{\delta\Pi}{\pi G\rho}
\nonumber
\\-5\overline{\eta}\left[\overline{\lambda}V_{11}+\frac{\overline{2\Omega}f}{1+\alpha^2}V_{12}\right]
\nonumber
\\-\frac{8\alpha\beta(\alpha^2+\beta^2)}{75}\overline{t}_c^5\left[\frac{1}{3}\overline{\lambda}^5(2V_{11}-V_{22}-V_{33})-\phi_1(\overline{\lambda},\overline{\Omega})(V_{11}-V_{22})+\left(16\overline{\Omega}^5\frac{\alpha^2-1}{\alpha^2+\beta^2}-2\phi_2(\overline{\lambda},\overline{\Omega})\right)V_{12}\right],
\label{eq:CharacteristicEqn11R}
\\
\left[\frac{1}{2}\overline{\lambda}^2-\frac{\alpha^2\overline{\Omega}^2f^2}{(1+\alpha^2)^2}-\overline{\Omega}^2\right]V_{22}-\frac{2\alpha^2\overline{\Omega}\overline{\lambda}f}{1+\alpha^2}V_{2;1}+2\overline{\lambda}\overline{\Omega} V_{1;2}-\frac{\overline{\Omega}^2f}{1+\alpha^2}(V_{22}+\alpha^2V_{11})=-\left[2\overline{B}_{22}V_{22}-\alpha^2\sum^3_{\ell=1}\overline{A}_{2\ell}V_{\ell\ell}\right]+\frac{\delta\Pi}{\pi G\rho}
\nonumber
\\-5\overline{\eta}\left[\overline{\lambda}\frac{V_{22}}{\alpha^2}-\frac{2\overline{\Omega}f}{1+\alpha^2}V_{12}\right]
\nonumber
\\-\frac{8\alpha\beta(1+\beta^2)}{75}\overline{t}_c^5\left[\frac{1}{3}\overline{\lambda}^5(2V_{22}-V_{11}-V_{33})-\phi_1(\overline{\lambda},\overline{\Omega})(V_{22}-V_{11})+\left(16\overline{\Omega}^5\frac{\alpha^2-1}{1+\beta^2}+2\phi_2(\overline{\lambda},\overline{\Omega})\right)V_{12}\right],
\label{eq:CharacteristicEqn22R}
\\
\overline{\lambda}^2V_{1;2}-\frac{\alpha^2\overline{\Omega}\overline{\lambda}f}{1+\alpha^2}V_{11}-\overline{\lambda}\overline{\Omega} V_{22}=-5\overline{\eta}\left[\overline{\lambda}\left(\frac{V_{1;2}}{\alpha^2}+V_{2;1}\right)-\frac{\overline{\Omega}f}{2}\left(V_{11}-\frac{V_{22}}{\alpha^2}\right)\right]
\nonumber
\\-\frac{8\alpha\beta(1+\beta^2)}{75}\overline{t}_c^5\left[16\overline{\Omega}^5\frac{\alpha^2-1}{1+\beta^2}V_{22}+(\overline{\lambda}^5-2\phi_1(\overline{\lambda},\overline{\Omega}))V_{12}+\phi_2(\overline{\lambda},\overline{\Omega})(V_{11}-V_{22})\right],
\label{eq:CharacteristicEqn12R}
\\
\overline{\lambda}^2V_{2;1}+\frac{\overline{\Omega}\overline{\lambda}f}{1+\alpha^2}V_{22}+\overline{\lambda}\overline{\Omega} V_{11}=-5\overline{\eta}\left[\overline{\lambda}\left(\frac{V_{1;2}}{\alpha^2}+V_{2;1}\right)-\frac{\overline{\Omega}f}{2}\left(V_{11}-\frac{V_{22}}{\alpha^2}\right)\right]
\nonumber
\\-\frac{8\alpha\beta(\alpha^2+\beta^2)}{75}\overline{t}_c^5\left[16\overline{\Omega}^5\frac{\alpha^2-1}{\alpha^2+\beta^2}V_{11}+(\overline{\lambda}^5-2\phi_1(\overline{\lambda},\overline{\Omega}))V_{12}+\phi_2(\overline{\lambda},\overline{\Omega})(V_{11}-V_{22})\right],
\label{eq:CharacteristicEqn21R}
\end{align}
for $V_{i;j}$ even in index 3, and for those odd in index 3

\begin{align}
\overline{\lambda}^2 V_{1;3}-2\overline{\lambda}\overline{\Omega} V_{2;3}-\frac{\alpha^2\overline{\Omega}^2f^2}{(1+\alpha^2)^2}V_{3;1}- \frac{2\alpha^2\overline{\Omega}^2f}{1+\alpha^2}V_{3;1}={}&
(\overline{\Omega}^2-2\overline{B}_{13})V_{13}-5\overline{\lambda}\overline{\eta}\left(\frac{V_{1;3}}{\beta^2}+V_{3;1}\right)
\nonumber
\\
{}&-\frac{8\alpha\beta(1+\alpha^2)}{75}\overline{t}_c^5\overline{\lambda}^5V_{13}-\frac{128\alpha\beta(\alpha^2-1)}{75}\overline{t}_c^5\overline{\Omega}^5V_{23},
\label{eq:CharacteristicEqn13R}
\\
\overline{\lambda}^2 V_{2;3}+2\overline{\lambda}\overline{\Omega} V_{1;3}-\frac{\alpha^2\overline{\Omega}^2f^2}{(1+\alpha^2)^2}V_{3;2}-\frac{2\overline{\Omega}^2f}{1+\alpha^2}V_{3;2}={}&
(\overline{\Omega}^2-2\overline{B}_{23})V_{23}-5\overline{\eta}\left[\overline{\lambda}\left(\frac{V_{2;3}}{\beta^2}+\frac{V_{3;2}}{\alpha^2}\right)-\frac{\overline{\Omega}f}{1+\alpha^2}\left(V_{3;1}+\frac{\alpha^2}{\beta^2}V_{1;3}\right)\right]
\nonumber
\\
{}&-\frac{8\alpha\beta(1+\alpha^2)}{75}\overline{t}_c^5\overline{\lambda}^5V_{23}-\frac{128\alpha\beta(\alpha^2-1)}{75}\overline{t}_c^5\overline{\Omega}^5V_{13},
\label{eq:CharacteristicEqn23R}
\\
\overline{\lambda}^2 V_{3;1}+\frac{2\overline{\Omega}\overline{\lambda}f}{1+\alpha^2}V_{3;2}-\frac{\alpha^2\overline{\Omega}^2f^2}{(1+\alpha^2)^2}V_{3;1}={}&
-2\overline{B}_{13}V_{13}-5\overline{\lambda}\overline{\eta}\left(V_{3;1}+\frac{V_{1;3}}{\beta^2}\right)
\nonumber
\\
{}&-\frac{8\alpha\beta(\alpha^2+\beta^2)}{75}\overline{t}_c^5\left([\overline{\lambda}^5-\phi_1(\overline{\lambda},\overline{\Omega})]V_{13}-\phi_2(\overline{\lambda},\overline{\Omega})V_{23}\right),
\label{eq:CharacteristicEqn31R}
\\
\overline{\lambda}^2 V_{3;2}-\frac{2\alpha^2\overline{\Omega}\overline{\lambda}f}{1+\alpha^2}V_{3;1}-\frac{\alpha^2\overline{\Omega}^2f^2}{(1+\alpha^2)^2}V_{3;2}={}&
-2\overline{B}_{23}V_{23}-5\overline{\eta}\left[\overline{\lambda}\left(\frac{V_{3;2}}{\alpha^2}+\frac{V_{2;3}}{\beta^2}\right)-\frac{\overline{\Omega}f}{1+\alpha^2}\left(V_{3;1}+\frac{\alpha^2}{\beta^2}V_{1;3}\right)\right]
\nonumber
\\
{}&-\frac{8\alpha\beta(1+\beta^2)}{75}\overline{t}_c^5\left([\overline{\lambda}^5-\phi_1(\overline{\lambda},\overline{\Omega})]V_{23}+\phi_2(\overline{\lambda},\overline{\Omega})V_{13}\right),
\label{eq:CharacteristicEqn32R}
\end{align}
where  $\overline{A}_{ij}$ and $\overline{B}_{ij}$ are 
obtained from the index symbols 
${A}_{ij}$ and ${B}_{ij}$ via the replacement $a_1\rightarrow1$, $a_2\rightarrow\alpha=a_2/a_1$, $a_3\rightarrow\beta=a_3/a_1$.
Eqs.~\eqref{eq:CharacteristicEqn13R}--\eqref{eq:CharacteristicEqn23R} can be simplified using Eq. (36)--(37) of Chap. 7 of EFE:
\begin{align}
    \Omega^2-Q_{12}Q_{21}+2Q_{21}\Omega={}&2\frac{a_1^2-a_3^2}{a_1^2}\pi G\rho B_{13},
    \\
    \Omega^2-Q_{12}Q_{21}-2Q_{12}\Omega={}&2\frac{a_2^2-a_3^2}{a_2^2}\pi G\rho B_{23}.
\end{align}
to obtain 
\begin{align}
\overline{\lambda}^2 V_{1;3}-2\overline{\lambda}\overline{\Omega} V_{2;3}+2\beta^2\overline{B}_{13}V_{3;1}={}&
(\overline{\Omega}^2-2\overline{B}_{13})V_{1;3}-5\overline{\lambda}\overline{\eta}\left(\frac{V_{1;3}}{\beta^2}+V_{3;1}\right)-\frac{8\alpha\beta(1+\alpha^2)}{75}\overline{t}_c^5\overline{\lambda}^5V_{13}
\nonumber
\\
{}&-\frac{128\alpha\beta(\alpha^2-1)}{75}\overline{t}_c^5\overline{\Omega}^5V_{23},
\label{eq:CharacteristicEqn13R2}
\\
\overline{\lambda}^2 V_{2;3}+2\overline{\lambda}\overline{\Omega} V_{1;3}+\frac{2\beta^2}{\alpha^2}\overline{B}_{23}V_{3;2}={}&
(\overline{\Omega}^2-2\overline{B}_{23})V_{2;3}-5\overline{\lambda}\overline{\eta}\left(\frac{V_{2;3}}{\beta^2}+\frac{V_{3;2}}{\alpha^2}\right)-\frac{8\alpha\beta(1+\alpha^2)}{75}\overline{t}_c^5\overline{\lambda}^5V_{23}
\nonumber
\\
{}&-\frac{128\alpha\beta(\alpha^2-1)}{75}\overline{t}_c^5\overline{\Omega}^5V_{13}.
\label{eq:CharacteristicEqn23R2}
\end{align}

\section{Gravitational radiation back-reaction term in virial equation 2: Full 2.5-post-Newtonian form}
\label{app:GWBackReaction2}

Our derivation is similar, but more general, than that of~\citet{Chandrasekhar1970a}.
In the rest frame of the center of mass, the radiation reaction terms in the equations of motion are~\citep{Chandrasekhar1970}
\begin{equation}
f_{\text{GW}}^{a}=\frac{1}{c}T^{a j}_{\ \ ;j}=\frac{1}{c^5}\left[-\rho Q_{00}^{(5)}\frac{\text{d}v_{a}}{\text{d}t}-\frac{1}{2}\rho v_{a}\frac{\text{d}Q_{00}^{(5)}}{\text{d}t}-\rho\frac{\text{d}}{\text{d}t}\left(v_{i}Q_{i a}^{(5)}\right)-\frac{1}{2}\rho Q_{ij}^{(5)}\frac{\text{d}\mathfrak{B}_{ij}}{\text{d}x_{a}}+\frac{1}{5}\rho x_{a}G\frac{\text{d}^5I_{ii}}{\text{d}t^5}-\frac{3}{5}\rho x_{i}G\frac{\text{d}^5I_{i a}}{\text{d}t^5}\right],
\label{eq:GWBackReaction}
\end{equation}
where the Einstein summation convention is used. The Latin indices that are not explicitly summed over run over the spatial indices 1,2,3. $v_{a}$ is the velocity as measured in the inertial frame. $Q_{00}^{(5)}$ and  $Q_{ab}^{(5)}$ are given by
\begin{align}
    Q_{00}^{(5)}={}&\frac{4}{3}G\frac{\text{d}^3I_{ii}}{\text{d}t^3},
    \\
     Q_{ab}^{(5)}={}&2G\frac{\text{d}^3I_{ab}}{\text{d}t^3}-\frac{2}{3}G\delta_{ab}\frac{\text{d}^3I_{ii}}{\text{d}t^3},
\end{align}
where
$I_{ab}$ is the moment of inertia as measured in the inertial frame (although~\citet{Chandrasekhar1970a} does not explicitly state this) and $\mathfrak{B}_{ij}$ is defined in Eq.~(\ref{eq:BijSymbol}).

In the frame rotating with the star about the $x_3$ axis, which is the same in both rotating and inertial frames, the moment of inertia tensor $I_{ab}^{(r)}$ is constant and diagonal
\begin{equation}
    I^{(r)}_{ab}=\int_{\mathcal{V}}\text{d}^3x\rho x_{a}x_{b} = \delta_{ab}I_{aa}.
    \label{eq:MoIRotatingFrame}
\end{equation}
The time derivatives of the moment of inertia tensor in the inertial frame, $I^{(i)}_{ab}$, are related to those of $I^{(r)}_{ab}$ by Eq.~(14) of~\citet{Chandrasekhar1970a}:
\begin{equation}
\frac{\text{d}^nI^{(i)}_{ab}}{\text{d}t^n}=\sum^{n}_{m=0}\sum^{m}_{p=0}C^{n}_{m}C^{m}_{p}(-1)^{p}[(\boldsymbol{\Omega}^*)^p]_{abc}\frac{\text{d}^{n-m}I^{(r)}_{ck}}{\text{d}t^{n-m}}
[(\boldsymbol{\Omega}^*)^{m-p}]_{k b}=\Omega^n\sum^n_{p=0}C^n_p(-1)^p[\boldsymbol{\sigma}^p]_{ac}I^{(r)}_{ck}[\boldsymbol{\sigma}^{n-p}]_{k b},
\label{eq:MoIInertial}
\end{equation}
where $\boldsymbol{\sigma}_{ab}$ is
\begin{equation}
\boldsymbol{\sigma}_{ab}=\left(\begin{array}{ccc} 0 & 1 & 0 \\ -1 & 0 & 0 \\ 0 & 0 & 0 \end{array} \right).
\end{equation}
From Eq.~(\ref{eq:MoIInertial}), one finds (\citet{Chandrasekhar1970a} Eq.~(18))
\begin{equation}
   \frac{\text{d}^{2n+1}I^{(i)}_{ab}}{\text{d}t^{2n+1}} = (-1)^n2^{2n}\Omega^{2n+1}(I_{11}-I_{22})\left(\begin{array}{ccc} 0 & 1 & 0 \\ 1 & 0 & 0 \\ 0 & 0 & 0 \end{array} \right).
\end{equation}
So $Q_{00}^{(5)}=0$ and $\text{d}^5I^{(i)}_{ab}/\text{d}t^5=0$ are still true as in~\citet{Chandrasekhar1970a}, but their perturbations are not necessarily zero.

We are also going to be concerned with the perturbed form of this when we compute the contribution of gravitational radiation back-reaction to the second-order virial equation. Noting that
\begin{equation}
    \delta I^{(r)}_{ab}=\int_{\mathcal{V}}\text{d}^3x\rho\left(\xi_{a}x_{b}+x_{a}\xi_{b}\right),
\end{equation}
and assuming (as in the main text) $\xi_{i}(\mathbf{x},t)=e^{\lambda t}\xi_{i}(\mathbf{x})$, we have 
\begin{equation}
    \frac{\text{d}^{n-m}I^{(r)}_{ab}}{\text{d}t^{n-m}}=\lambda^{n-m}V_{ab},
\end{equation}
and hence the perturbation of the time derivatives of the inertial frame moment of inertia are
\begin{equation}
    \delta I^{(n)}_{ab}\equiv \frac{\text{d}^n\delta I^{(i)}_{ab}}{\text{d}t^n}=\sum^{n}_{m=0}\sum^{m}_{p=0}\lambda^{n-m}\Omega^m C^{n}_{m}C^{m}_{p}(-1)^{p}[\boldsymbol{\sigma}^p]_{ac}V_{ck}[\boldsymbol{\sigma}^{m-p}]_{k b}.
\end{equation}
\citet{Chandrasekhar1970a} often uses the abbreviation $\delta I^{(n)}_{ab}$, and so do we to be able to compare to his results for  Maclaurin spheroid. The forms of this tensor for $n=3,4,5$ are given by~\citet{Chandrasekhar1970a} Eq.~(26)--(28).

We also need the background velocity $v_{a}$ and acceleration $\text{d}v_{a}/\text{d}t$, which in the Riemann S-type ellipsoid case are
\begin{align}
    v_{a}{}&=u_{a}+\varepsilon_{abi}\Omega_{b}x_{i}\equiv\Omega\tilde{S}_{ab}x_{b},
    \\
    \frac{\text{d}v_{a}}{\text{d}t}{}&=\frac{\text{d}u_{a}}{\text{d}t}+\varepsilon_{abi}\Omega_{b}\varepsilon_{ic j}\Omega_{c}x_{j}+2\varepsilon_{abi}\Omega_{b}u_{i}=-\Omega^2\tilde{R}_{a}x_{a},
\end{align}
where $u_{a}$ is the background velocity in the rotating frame defined in Eq.~\eqref{eq:u1}--\eqref{eq:u3}. This can be compared to Eq.~(31) in \citet{Chandrasekhar1970a}  where in the Maclaurin ellipsoid case the vorticity is zero so $u_{a} = 0$. The nonzero $\tilde{S}_{ab}$ are
\begin{align}
    \tilde{S}_{12}=-1-\frac{a_1^2f}{a_1^2+a_2^2}, \qquad \tilde{S}_{21}=1+\frac{a_2^2f}{a_1^2+a_2^2},
\end{align}
and the nonzero $\tilde{R}_{a}$ are
\begin{align}
    \tilde{R}_{1}={}&1+\frac{a_2^2f}{a_1^2+a_2^2}+\frac{a_1^2a_2^2f^2}{(a_1^2+a_2^2)^2}, \qquad \tilde{R}_{2}=1+\frac{a_1^2f}{a_1^2+a_2^2}+\frac{a_1^2a_2^2f^2}{(a_1^2+a_2^2)^2},
\end{align}
and $v_{3}=0=\text{d}v_{3}/\text{d}t$.

Taking the perturbation of Eq.~(\ref{eq:GWBackReaction}) and using that $Q_{00}^{(5)}=0$, $\text{d}^5I^{(i)}_{ab}/\text{d}t^5=0$, we find
\begin{align}
\delta f_{\text{GW}}^{a}={}&\frac{1}{c^5}\delta\left[-\rho\frac{\text{d}v_{i}}{\text{d}t}Q_{ia}^{(5)}-\rho v_{i}\frac{\text{d}Q_{i a}^{(5)}}{\text{d}t}-\frac{1}{2}\rho Q_{ij}^{(5)}\frac{\text{d}\mathfrak{B}_{ij}}{\text{d}x_{a}}-\frac{3}{5}\rho x_{i}G\frac{\text{d}^5I_{i a}}{\text{d}t^5}\right]
\nonumber
\\
={}&\frac{1}{c^5}\Bigg[-\rho\frac{\text{d}\delta v_{i}}{\text{d}t}Q_{ia}^{(5)}-\rho \delta v_{i}\frac{\text{d}Q_{i a}^{(5)}}{\text{d}t}-\rho\frac{\text{d}v_{i}}{\text{d}t}\delta Q_{ia}^{(5)}-\rho v_{i}\frac{\text{d}\delta Q_{i a}^{(5)}}{\text{d}t}-\frac{1}{2}\rho \delta Q_{ij}^{(5)}\frac{\text{d}\mathfrak{B}_{ij}}{\text{d}x_{a}}-\frac{1}{2}\rho Q_{ij}^{(5)}\frac{\text{d}\delta\mathfrak{B}_{ij}}{\text{d}x_{a}}
\nonumber
\\
{}&\qquad\qquad-\frac{3}{5}\rho \delta x_{i}G\frac{\text{d}^5I_{i a}}{\text{d}t^5}-\frac{3}{5}\rho x_{i}G\frac{\text{d}^5\delta I_{i a}}{\text{d}t^5}-\rho\delta Q_{00}^{(5)}\frac{\text{d}v_{a}}{\text{d}t}-\frac{1}{2}\rho v_{a}\frac{\text{d}\delta Q_{00}^{(5)}}{\text{d}t}+\frac{1}{5}\rho x_{a}G\frac{\text{d}^5\delta I_{ii}}{\text{d}t^5}\Bigg].
\label{eq:GWBackReactionPerturbation}
\end{align}
The contribution to the second-order virial equation, denoted $\delta\mathcal{G}_{ij}$ in Eq.~(\ref{eq:CharacteristicEquation}), is
\begin{equation}
    \delta\mathcal{G}_{ab}=\int_{\mathcal{V}}\text{d}^3xx_{b}\delta f_{a}^{\text{GW}}.
\end{equation}
Using that $\delta x_{i}=\xi_{i}$, the eight terms on the right-hand side of this are thus (labeled with a superscript number in brackets)
\begin{equation}
    \delta\mathcal{G}_{ab}^{(1)}\equiv -\frac{1}{c^5}\int_{\mathcal{V}}\text{d}^3x\left[\rho x_{b}\frac{\text{d}\delta v_{i}}{\text{d}t}Q_{ia}^{(5)}\right]=\frac{\Omega^2}{c^5}\int_{\mathcal{V}}\text{d}^3x\left[\rho x_{b}\tilde{R}_{i}\xi_{i}Q_{ia}^{(5)}\right]=-\frac{8G\Omega^5}{c^5}(I_{11}-I_{22})\left(\tilde{R}_1\delta^2_{a}V_{1;b}+\tilde{R}_2\delta^1_{a}V_{2;b}\right)
\end{equation}

\begin{align}
    \delta\mathcal{G}_{ab}^{(2)}\equiv{}& -\frac{1}{c^5}\int_{\mathcal{V}}\text{d}^3x\left[\rho x_{b}\delta v_{i}\frac{\text{d} Q_{ia}^{(5)}}{\text{d}t}\right]=-\frac{\Omega}{c^5}\int_{\mathcal{V}}\text{d}^3x\left[\rho x_{b}\left(\tilde{S}_{21}\delta_a^2\xi_1+\tilde{S}_{12}\delta^1_a\xi_2\right)\frac{\text{d} Q_{ia}^{(5)}}{\text{d}t}\right]=
    \nonumber
    \\
    ={}&-\frac{16G\Omega^5}{c^5}(I_{11}-I_{22})\left(\tilde{S}_{12}\delta_{a}^1V_{2;b}-\tilde{S}_{21}\delta^2_{a}V_{1;b}\right),
\end{align}
where we used Eq.~(17) of \cite{Chandrasekhar1970a},
\begin{equation}
   \frac{\text{d}^{2n}I^{(i)}_{ab}}{\text{d}t^{2n}} = (-1)^n2^{2n-1}\Omega^{2n}(I_{11}-I_{22})\left(\begin{array}{ccc} 1 & 0 & 0 \\ 0 & -1 & 0 \\ 0 & 0 & 0 \end{array} \right)_{ab},
\end{equation}

\begin{align}
    \delta\mathcal{G}_{ab}^{(3)}{}&\equiv -\frac{1}{c^5}\int_{\mathcal{V}}\text{d}^3x\left[\rho x_{b}\frac{\text{d}v_{i}}{\text{d}t}\delta Q_{ia}^{(5)}\right]=\frac{2G\Omega^2}{c^5}\int_{\mathcal{V}}\text{d}^3x\left[\rho x_{b}\left(\tilde{R}_1x_1\delta^1_{i}+\tilde{R}_2x_2\delta^2_{i}\right)\delta Q_{ia}^{(3)}\right]
    \nonumber
    \\
    {}&=\frac{2G\Omega^2}{c^5}\left[I_{11}\tilde{R}_1\delta_{b}^1\left(\delta I_{1a}^{(3)}-\frac{1}{3}\delta_{1a}\delta I^{(3)}_{jj}\right) + I_{22}\tilde{R}_{2}\delta_{b}^2\left(\delta I_{2a}^{(3)}-\frac{1}{3}\delta_{2a}\delta I^{(3)}_{jj}\right)\right],
\end{align}
where we  used that the moment of inertia tensor in the rotating frame is diagonal i.e. Eq.~(\ref{eq:MoIRotatingFrame}),
\begin{equation}
    \delta\mathcal{G}_{ab}^{(4)}\equiv -\frac{1}{c^5}\int_{\mathcal{V}}\text{d}^3x\left[\rho x_{b}v_{i}\frac{\text{d}\delta Q_{i a}^{(5)}}{\text{d}t}\right]=-\frac{2G\Omega}{c^5}\left[I_{11}\tilde{S}_{21}\delta^1_{b}\left(\delta I^{(4)}_{2a}-\frac{1}{3}\delta_{2a}\delta I^{(4)}_{jj}\right)+I_{22}\tilde{S}_{12}\delta^2_{b} \left(\delta I^{(4)}_{1a}-\frac{1}{3}\delta_{1a}\delta I^{(4)}_{jj}\right)\right],
\end{equation}

\begin{equation}
     \delta\mathcal{G}_{ab}^{(5)}\equiv-\frac{1}{2c^5}\int_{\mathcal{V}}\text{d}^3x\rho x_{b}\delta Q_{ij}^{(5)}\frac{\text{d}\mathfrak{B}_{ij}}{\text{d}x_{a}}=-\frac{4G(\pi G\rho)}{c^5}\left[B_{ab}I_{bb}\left(\delta I^{(3)}_{ab}-\frac{1}{3}\delta_{ab}\delta I^{(3)}_{jj}\right)-\frac{1}{3}a_{i}^2\delta_{ab}I_{aa}A_{ia}\delta I^{(3)}_{ii}\right],
\end{equation}
where we used
\begin{equation}
    \frac{\text{d}\mathfrak{B}_{ij}}{\text{d}x_{a}}=\pi G\rho\left[2B_{ij}(x_{j}\delta_{i}^{a}+x_{i}\delta^{a}_{j})-2a_{i}^2\delta_{ij}A_{ia}x_{a}\right],
\end{equation}

\begin{equation}
     \delta\mathcal{G}_{ab}^{(6)}\equiv-\frac{1}{2c^5}\int_{\mathcal{V}}\text{d}^3x\rho x_{b}Q_{ij}^{(5)}\frac{\text{d}\delta\mathfrak{B}_{ij}}{\text{d}x_{a}}=\frac{16G(\pi G\rho)\Omega^3}{c^5}B_{12}(I_{11}-I_{22})\left(\delta^2_{a}V_{1;b}+\delta^1_{a}V_{2;b}\right),
\end{equation}
where we used
\begin{equation}
    \frac{\text{d}\delta\mathfrak{B}_{ij}}{\text{d}x_{a}}=\pi G\rho\left[2B_{ij}(\xi_{j}\delta_{i}^{a}+\xi_{i}\delta^{a}_{j})-2a_{i}^2\delta_{ij}A_{ia}\xi_{a}\right],
\end{equation}

\begin{equation}
     \delta\mathcal{G}_{ab}^{(7)}\equiv-\frac{3}{5c^5}\int_{\mathcal{V}}\text{d}^3x\rho x_{b}\delta x_{i}G\frac{\text{d}^5I_{i a}}{\text{d}t^5}=-\frac{48G\Omega^5}{5c^5}B_{12}(I_{11}-I_{22})\left(\delta^2_{a}V_{1;b}+\delta^1_{a}V_{2;b}\right),
\end{equation}

\begin{equation}
     \delta\mathcal{G}_{ab}^{(8)}\equiv-\frac{3}{5c^5}\int_{\mathcal{V}}\text{d}^3x\rho x_{b}x_{i}G\frac{\text{d}^5\delta I_{i a}}{\text{d}t^5}=-\frac{3G}{5c^5}I_{bb}\delta I^{(5)}_{ba},
\end{equation}

\begin{equation}
     \delta\mathcal{G}_{ab}^{(9)}\equiv-\frac{1}{c^5}\int_{\mathcal{V}}\text{d}^3x\rho x_{b}\delta Q_{00}^{(5)}\frac{\text{d}v_{a}}{\text{d}t}=\frac{4G\Omega^2}{c^5}\left(I_{11}\tilde{R}_1\delta_{a}^1\delta_{b}^1+I_{22}\tilde{R}_2\delta_{a}^2\delta_{b}^2\right)\delta I^{(3)}_{ii},
\end{equation}

\begin{equation}
     \delta\mathcal{G}_{ab}^{(10)}\equiv-\frac{1}{2c^5}\int_{\mathcal{V}}\text{d}^3x\rho x_{b}v_{a}\frac{\text{d}\delta Q_{00}^{(5)}}{\text{d}t}=-\frac{2G\Omega}{3c^5}\left(I_{11}\tilde{S}_{21}\delta_{a}^2\delta_{b}^1+I_{22}\tilde{S}_{12}\delta_{a}^1\delta_{b}^2\right)\delta I^{(4)}_{ii},
\end{equation}

\begin{equation}
     \delta\mathcal{G}_{ab}^{(11)}\equiv\frac{1}{5c^5}\int_{\mathcal{V}}\text{d}^3x\rho x_{b}x_{a}G\frac{\text{d}^5\delta I_{i i}}{\text{d}t^5}=\frac{G}{5c^5}I_{bb}\delta_{ab}\delta 
     I^{(5)}_{ii}.
\end{equation}

In the case of Maclaurin spheroid  discussed in~\citet{Chandrasekhar1970a}, $a_1=a_2$ and so $I_{11}=I_{22}$, in which case many of the contributions to $\delta\mathcal{G}_{ab}$ are eliminated. If we also discard the $V_{ab}$ with either of the indices being equal 3, the remaining terms reproduce
 Eq.~(35)--(38) of \citet{Chandrasekhar1970a}.

\end{onecolumn}

\bsp	
\label{lastpage}
\end{document}